\documentclass[prl,superscriptaddress,showpacs,amsmath,amssymb]{revtex4-1}
\usepackage{graphicx}
\usepackage{indentfirst}
\usepackage{psfrag}
\usepackage{epsfig}
\usepackage{amsmath}
\usepackage{amssymb}
\usepackage{bm}

\begin{document}

%\begin{frontmatter}



\title{Composites of resonant dielectric rods: A test of their behavior
as metamaterial refractive elements}

\author{F.J. Valdivia-Valero$^{1}$ and M. Nieto-Vesperinas$^{1}$}

\address{$^{1}$Instituto de Ciencia de Materiales de Madrid,
C.S.I.C., Campus de Cantoblanco \\ 28049 Madrid, Spain}

\begin{abstract}
We report numerical experiments of optical wave propagation in
composites of high refractive index dielectric rods at frequencies
where their first electric and magnetic Mie resonances are excited.
The arrays of these particles have been extensively studied and
proposed as non-absorbing and isotropic metamaterials. We show that
negative refraction, observed in ordered particle arrays, is due to
diffraction and that an effective medium theory yields constitutive
parameters that do not reproduce the observations in these
composites, whose transmission also depends on the sample shape.
This is further confirmed by disordering the arrays, a case in which
large transmission losses appear due to extinction by resonant
scattering from the particles. Therefore, these composites although
little absorbing have large extinction due to scattering.
\end{abstract}
\maketitle
%\begin{keyword}

%\end{keyword}

%\end{frontmatter}

% \linenumbers

\section{Introduction}
Progress in metamaterial design, from microwaves to THz and optical
frequencies \cite{Soukoulis2002, Alu2005, Smith2005, Soukoulis2006,
Soukoulis2007}, based on the magnetic response of arrays of wires
and split ring resonators (SRR) and their modifications to shorter
electromagnetic waves, shows as essential limitations the existence
of anisotropy and large absorption losses \cite{Soukoulis2008}. More
recently, alternative structures based on Mie scattering by
dielectric spheres \cite{Soukoulis2008, Zhang2009_1, Mojahedi2005,
Mojahedi2006, Brongersma2007, Peng2007} of relative large refractive
index were extensively studied and proposed as models of lossless
composites at microwaves; these were also extended to semiconductor
cylinders and spheres, which were proven to possess similar resonant
characteristics in the infrared and visible regions \cite{Vynck2009,
Nieto2011}. Although initially these latter structures relied on
only the electric dipole or on multipole modes, later the
possibility of exciting the first Mie magnetic dipole resonance was
realized; which makes these particles equivalent from a fundamental
point of view to the microwave wire and SRR elements as far as their
electric and magnetic responses are concerned. Moreover, these
cylinders and spheres provide isotropy 2D and 3D in addition to
their resonance being subwavelength, both for the electric and
magnetic excited dipoles. The purpose of this paper is to
investigate the transmittance properties of media composed of these
dielectric particles and in particular, whether they behave as
uniform media at frequencies where their electric and magnetic
dipoles are excited. Therefore this work is a test and assessment of
whether the excitation of electric and magnetic resonances of high
index Mie spheres and cylinders constitutes an alternative with low
losses and high refractive transmittivity, to previously developed
metamaterial models.

In this connection, we emphasize that in spite of the exhaustive
studies on the possibilities of these resonant particles as
metamaterial building blocks in \cite{Mojahedi2005, Mojahedi2006,
Brongersma2007, Peng2007, Vynck2009, Nieto2011} and references
therein, no such a test has been carried out.

Although a unified view of the effective refractive index of
metamaterials made of ordered arrays of such particles, as well as
of photonic crystals (PC) producing negative refraction
\cite{Notomi2000} was established \cite{Zhang2010}, the
characterization of these composites as effective homogeneous media
to the propagating wave has generally employed the method of the
scattering (S) parameter by inversion of the complex transmittances
and reflectances \cite{Soukoulis2002}. Then, it remained the
question of whether the effective constitutive parameters
$\epsilon_{eff}$ and $\mu_{eff}$ derived from effective medium
homogenization procedures commonly employed \cite{Mojahedi2005,
Mojahedi2006, Peng2007, Nieto2011}, were the same as those obtained
by those methods that took into account the wave interaction with
the microstructure of the composite unit cell. The answer, recently
given \cite{Zhang2011}, is negative. {\it The linear dimension of
such unit cells, which is typically between $\lambda/10$ and
$\lambda/6$, is a much larger number than those of atom or molecule
arrangements in transparent dielectrics acting to the pass of
light}. In fact, the maximum lattice constant versus wavelength for
a metamaterial to behave as an uniform effective medium in negative
refraction experiments, was established in \cite{Zhang2008}.
Therefore, one may ask whether reducing absorption of the composite,
like with these Mie resonances of dielectric spheres and cylinders,
is sufficient to achieving an application as a metamaterial which
may be considered as a refractive element.

To this end, we shall see in this paper that, although contrary to
metal elements like those of earlier left - handed material (LHM)
designs, these dielectric cylinders and spheres grouped as "meta -
atoms{\lq\lq} of composites do not present absorption, their
scattering cross section is quite large. Therefore, {\it we will
demonstrate that this time the losses of light transmission come
from their large extinction cross section due to scattering, which
moreover, when random arrangements of these particles are also
addressed, yields a rather short transport mean free path, even if
the medium is homogenized, and hence the transmittivity of a
propagating beam through these media is low}.

As a 2-D equivalent to spheres, both ceramic \cite{Peng2007} and of
Silicon \cite{Vynck2009}, rod array composites were suggested to
exhibit metamaterial left-handed behaviour in the microwave and in
the visible to mid-infrared (IR) ranges, respectively, due to the
excitation of the cylinder magnetic Mie resonance. However, further
studies pointed out \cite{Zhang2009_2} the necessity of {\it more
research to short out whether the backward wave behavior inside a
periodic array of such Si cylinders is due to the band structure in
the diffraction regime \cite{Vanbesien2008} or to a pure left-handed
effect in the long wavelength range}, even though experiments of
microwaves in prisms of large permittivity ceramic rod arrays,
either ordered or random, suggested a left-handed behavior
\cite{Peng2007}.

The typical filling fraction $f$ of the studied dielectric cylinder
arrays \cite{Peng2007, Vynck2009, Nieto2011} is moderate $f\approx
0.30$, but the wavelength in vacuum to rod radius ratio: $\lambda/r$
was $61$ in e. g. the experiment of \cite{Peng2007}
($\lambda=41.64mm$), the lattice constant $a$ to $\lambda$ ratio
being 0.07, which is well below the aforementioned ratio $a/\lambda=
0.1$; whereas $\lambda/r= 9.8$ and $a/\lambda= 0.45$ in the model of
the ordered Si cylinders of \cite{Vynck2009} ($\lambda=1.55\mu m$).
On the other hand, the equivalent homogeneous media obtained from
Snell law had an index of negative refraction $n\simeq -0.6$ for the
PC slab of \cite{Vynck2009} and $n\simeq -1.08$ for the PC prism of
\cite{Peng2007}, ($\lambda_n\simeq 2.58\mu m$ for the case of
\cite{Vynck2009}, whereas $\lambda_n\simeq 38.5mm$ for
\cite{Peng2007}, $\lambda_n=\lambda/n$). None of these values is
maintained when one changes the sample geometry, as we shall prove
in this paper.

%From these numbers one infers that the long wave regime, and hence
%the effective medium theory (EMT) constitutive parameters should
%work better in e. g. the ceramic composite modeled in
%\cite{Peng2007} than in the Si arrangement of \cite{Vynck2009}.

In the homogenization procedure employed in the ceramic composite of
\cite{Peng2007}, the effective parameters obtained in the band of
left-handed behaviour have negative values both for the real part of
the magnetic permeability, $\mu^R_{eff}$, as for that of the
dielectric permittivity, $\epsilon^R_{eff}$, with $\mu^R_{eff}<<
\epsilon^R_{eff}$. However, near the resonance wavelength, the
imaginary parts $\epsilon^I_{eff}$ and $\mu^I_{eff}$ of both
constitutive parameters are non - negligible compared to those real
parts; {\it this conveys high extinction by the composite material,
which is not due to absorption losses as in previous models of LHMs
composed of metallic elements, but produced by scattering from the
high index particles of these metamaterials}. The same happens in
the near IR for the composite material of Si cylinders
\cite{Vynck2009} and \cite{Nieto2011}. This will be further studied
in this work.

In this paper we carry out 2 - D numerical experiments with the
finite element method (FEM) of propagation, of mid - IR waves,
through a composite medium made of dielectric rods either in
periodic or random positions. Nonetheless, for the sake of
comprehensiveness, we shall also address the related problem of
microwave propagation at larger scales. From the above discussion,
our aim is to assess for the first time to what extent these
structures may constitute a metamaterial model that overcomes the
losses \cite{Schultz2001, Nieto2003} of previous composites. This
conveys to discuss the validity of establishing effective
constitutive parameters. The results should closely predict
laboratory experimental observations because the calculations
involved are exact. Since in this model the size of the samples are
not huge compared to the wavelength \cite{Zhang2011, Zhang2008}, in
order to assess whether the propagation depends or not on this size
and on the shape of the sample, we employ two of such bodies for
observing transmission: a rectangular block, or thick slab, and a
prism, (i. e. in the 2-D calculations the latter being a triangle),
composed of either a periodic or a disordered array of rods in air.
Like in previous studies on these composites, the incident wave is
assumed to be linearly polarized with electric vector ${\bf E}$
along the cylinder axes.

\section{Sets of ordered and random rods as metamaterials}
\label{}

\subsection{Numerical procedure}
Maxwell equations are solved by using a finite element method (FE)
(FEMLAB of COMSOL, \mbox{http://www.comsol.com}). The calculation
domain is meshed with element growth rate: 1.55, meshing curvature
factor: 0.65, approximately. The geometrical resolution parameters
consist of 25 points per boundary segment to take into account
curved geometries in order to adapt the finite elements to the
geometry and optimize the convergence of the solution. The final
mesh contains about $10^{4}$ elements. To solve Helmholtz equation,
the UMFPACK direct is employed. The boundary conditions of the
simulation space are established both to keep the calculations from
undesired window reflections and to avoid possible geometrical
discontinuities. We then ensure that no inconsistencies due to
properties discontinuities of the objets under study appear, and
possible systematic errors are avoided.

In the 2 - D configuration, light depolarization is prevented by
launching linearly polarized light with propagation vector in the XY
- plane of the cylinder cross sections. Beam profiles are either,
rectangular (plane waves): ${\bf E_0}\exp(i({\bf k_i}\cdot {\bf
r_i}-\omega t))$, their widths being that of the simulation window,
or Gaussian: ${\bf E_0}\exp(-|{\bf R}-{\bf
R_0}|^2/2\sigma^2)\exp(i({\bf k_i}\cdot {\bf r}-\omega t))$, ${\bf
r}= ({\bf R}, z)$. ${\bf R}= (x, y)$ and ${\bf R_0}= (x_0, y_0)$ are
the transversal components of ${\bf r}$ and ${\bf r_0}$,
respectively, $\sigma$ is the standard deviation or beam waist, and
${\bf k_i}$ is the incident wave wavevector with ${\bf
|k_i|}=2\pi/\lambda$. For the wavelength $\lambda$, we shall address
values either in the microwave or IR regions. The direction of
propagation ${\bf k_i}$ of such beams is thus normally incident to
the OZ - axis of the infinite cylinders. The incident wave amplitude
is normalized to $|{\bf E_0}|= 1V/m$ (SI), which corresponds to a
magnitude of the time average energy flow $|<{\bf S}>|\approx
190W/m^{2}$. The criterium to choose the beam profile has been based
on the major response of the composite characteristics to analyze:
directionality of propagation in the media resulting from an
homogenization method, and extinction processes when the inner
structure of the rod distribution is considered. In the first case,
we employ rectangular incident beams; otherwise, we use Gaussian
beams.

The results are thus expressed in terms of either the electric
vector ${\bf E}({\bf R})$ which points along the cylinder OZ - axis,
the squared root of its time - averaged energy ${\bf |E(R)|}$, the
magnetic vector ${\bf H}({\bf R})= (H_x, H_y)$ or the time average
energy flow $<{\bf S}({\bf R})>$. These latter two vectors of course
being both transversal, namely, in the XY - plane of the images to
show next.

Finally, to classify the whispering gallery modes (WGM) associated
to the Mie resonances of the cylinders we will use the subscripts
\emph{(i, j)}, \emph{i} and \emph{j} standing for their angular
\emph{i - th} and radial \emph{j - th} orders, respectively.

\subsection{Electric and magnetic dipolar response of a dielectric cylinder in the microwave and mid - IR regimes}
\begin{figure}[htbp]
\begin{minipage}{.49\linewidth}
\centering
\includegraphics[width=6cm]{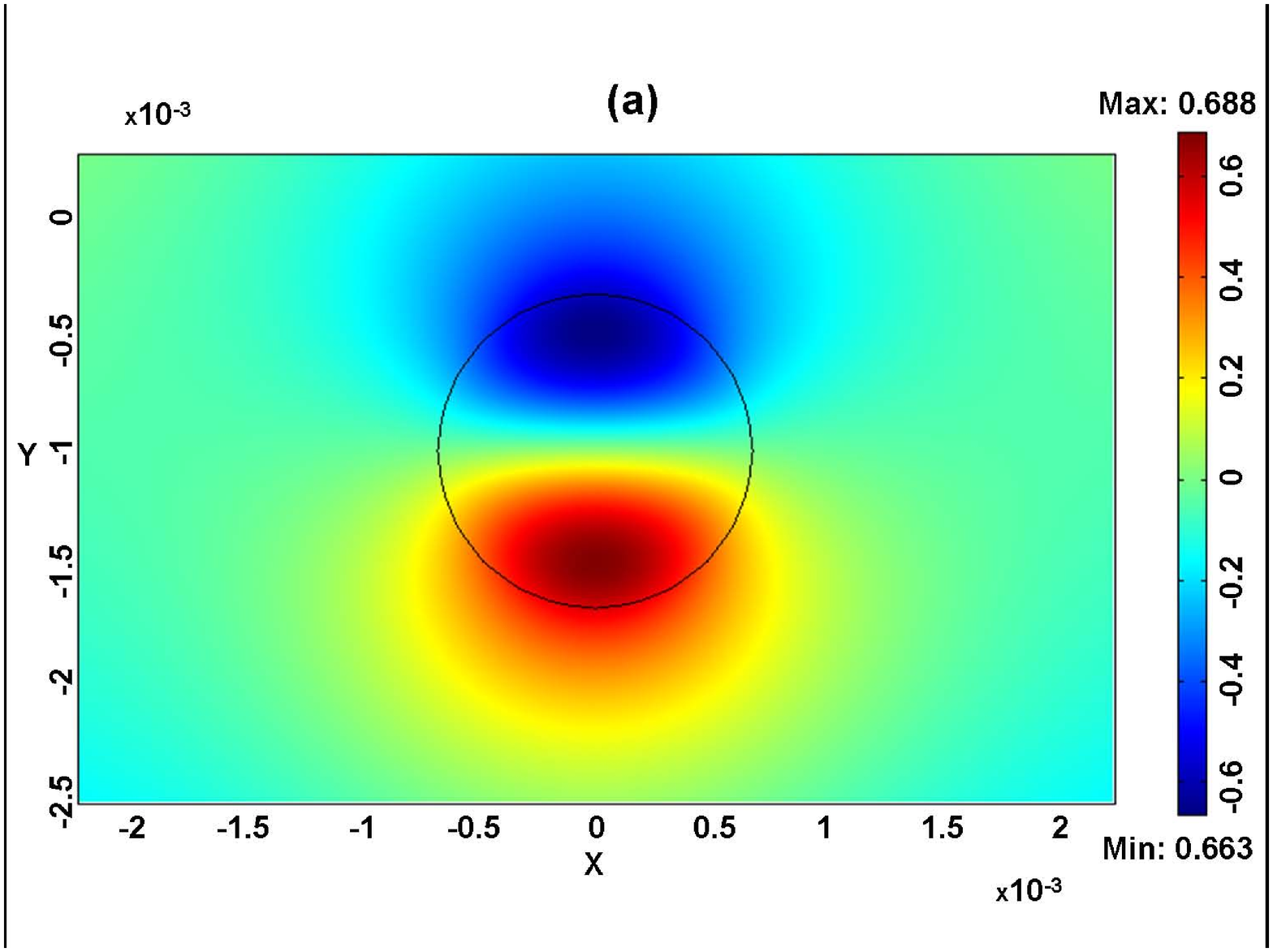}
\end{minipage}
\begin{minipage}{.49\linewidth}
\centering
\includegraphics[width=6cm]{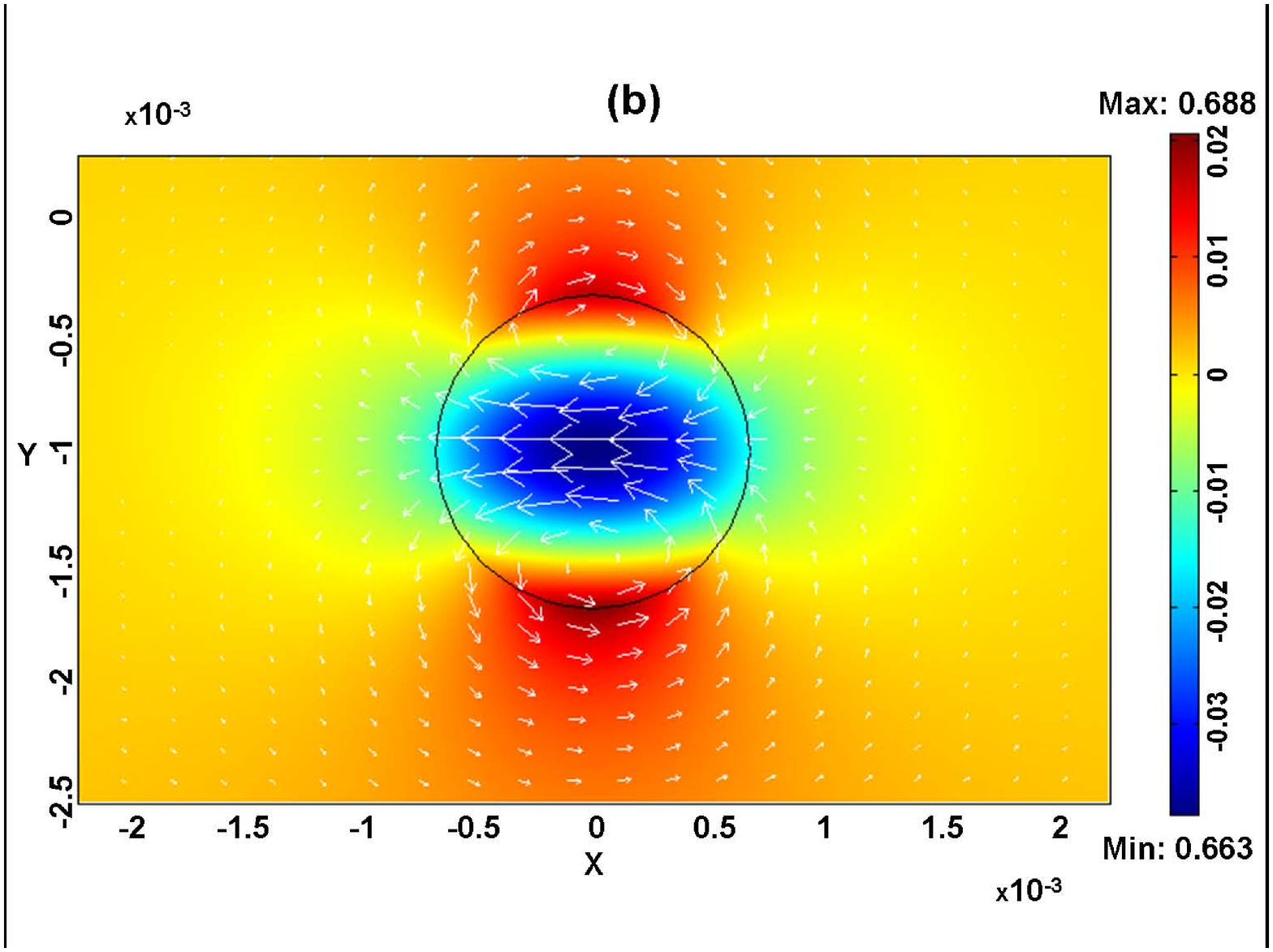}
\end{minipage}
\begin{minipage}{.98\linewidth}
\centering
\includegraphics[width=6cm]{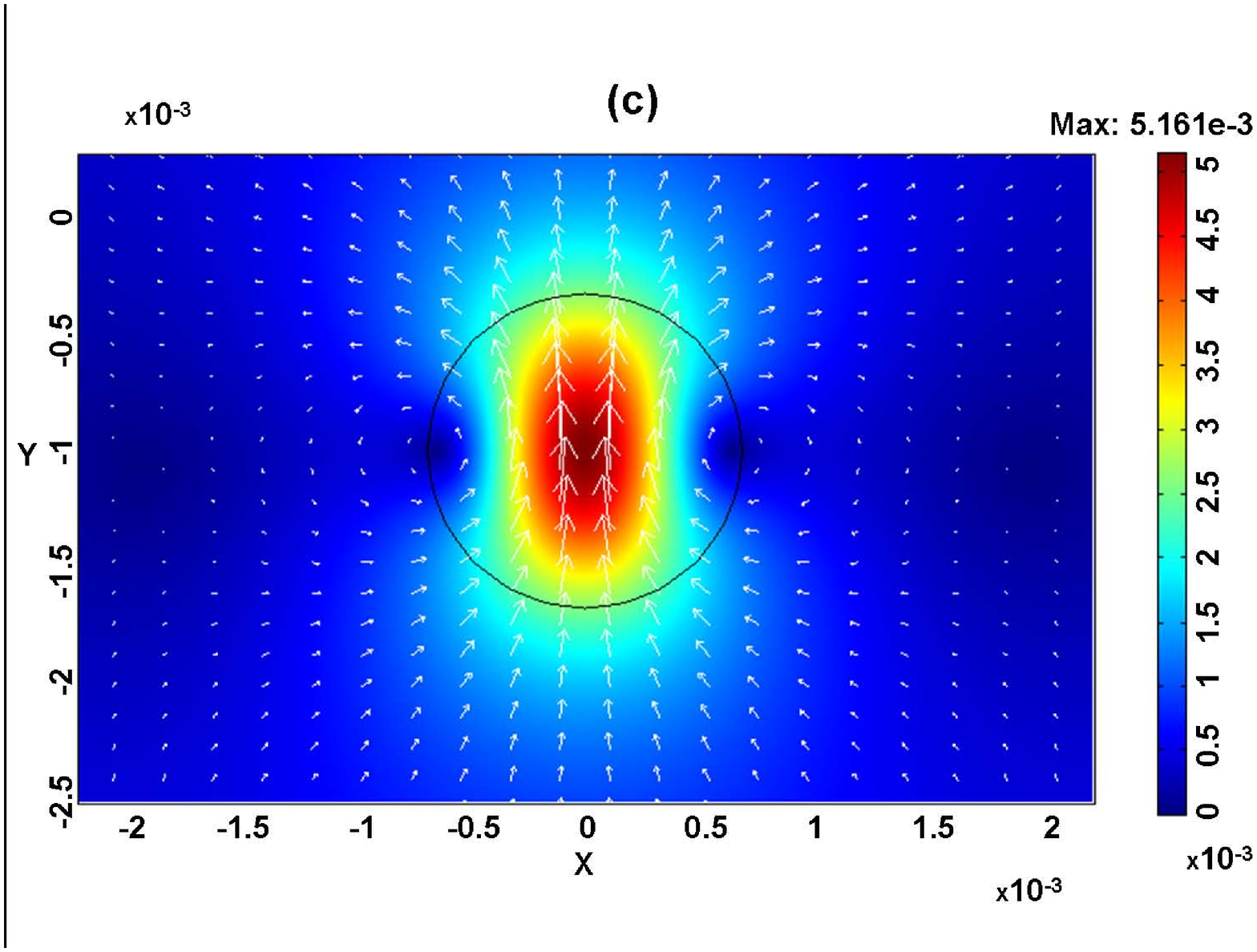}
\end{minipage}
\caption{(a) Electric field $E_z({\bf R})$ in a cylinder of BST
ceramic with a dielectric permittivity $\epsilon= 600$ and radius
$r= 0.68mm$. (b) Magnetic field ${\bf H(R)}$ (arrows) and its X -
component (colors). (c) Averaged energy flow ${\bf <S(R)>}$ (arrows)
and its norm (colors). In these figures, an s - polarized Gaussian
beam of amplitude $A= 1V/m$ and standard deviation $\sigma= 12mm$ at
$\lambda= 41.638mm$ is launched upwards (i. e. with ${\bf R_i}$
along the OY - axis), from below the cylinder, exciting its $WGM:
TM_{1,1}$.}
\end{figure}

{\noindent We first address the response of one single ceramic}
cylinder of $Ba_{0.5}Sr_{0.5}TiO_{3}$ (BST) to microwaves. Figures
1(a) - (c) show the electric, magnetic and time - averaged Poynting
vector distributions for linear polarized illumination with the
electric vector along the cylinder axis. Figure 1(a) exhibits
$E_z({\bf R})$, indicating two electric currents flowing along the
cylinder OZ - axis, one upwards and one downwards, respectively, and
centered near opposite sides of the rod periphery, which corresponds
to the {\bf E} - spatial distribution of the dipolar $WGM:
TM_{1,1}$. Figure 1(b) shows the magnetic vector lines in the XY -
plane, characterized by arrows circulating around these electric
currents, according to Ampere's law and behaves as that of a
magnetic dipole. The time - averaged energy flow, characterized by
the mean Poynting vector, is shown in Fig. 1(c) exhibiting an
interesting circulation around the equatorial extremes of the
cylinder section.

%\newpage

\begin{figure}[htbp]
\begin{minipage}{.49\linewidth}
\centering
\includegraphics[width=6cm]{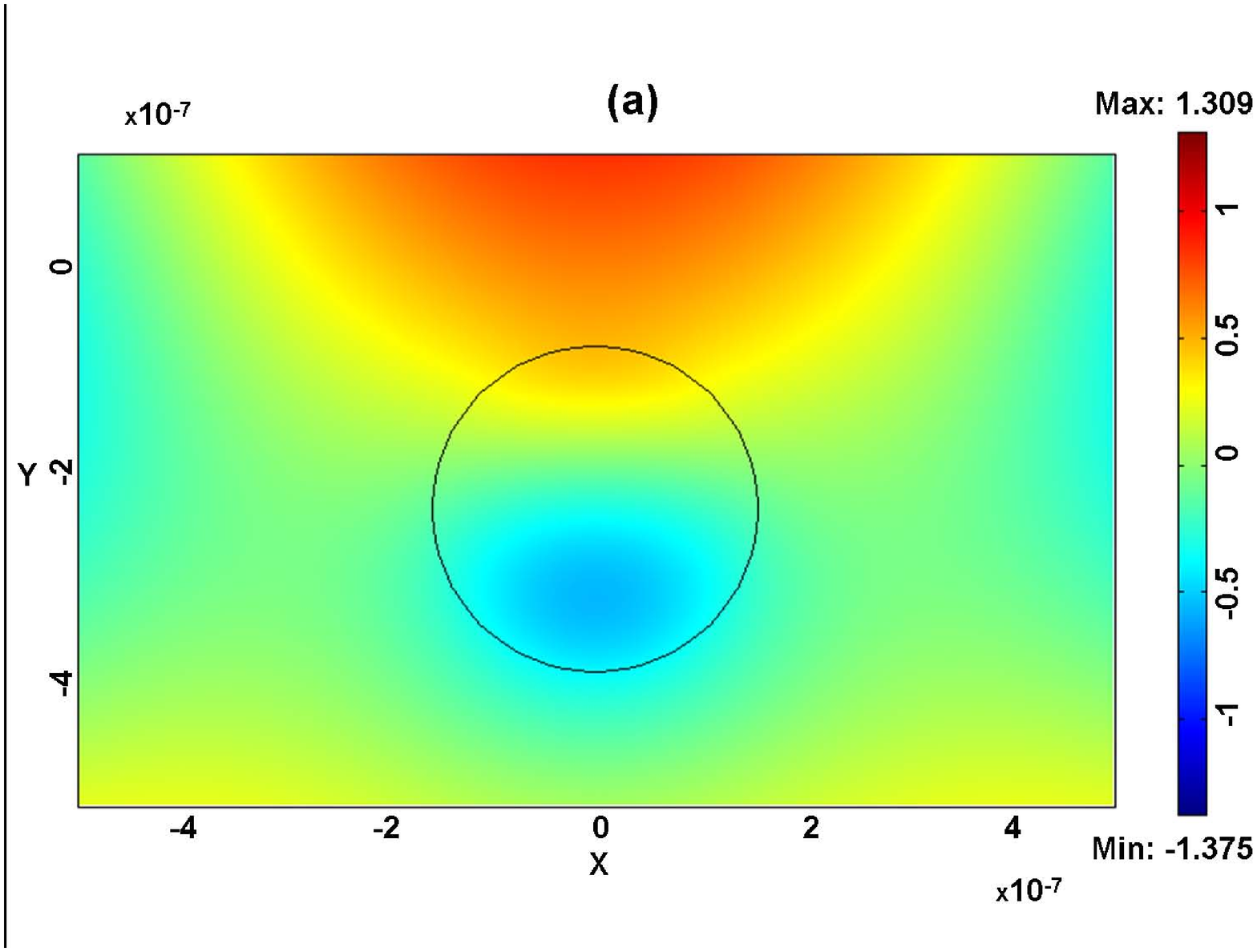}
\end{minipage}
\begin{minipage}{.49\linewidth} \centering
\includegraphics[width=6cm]{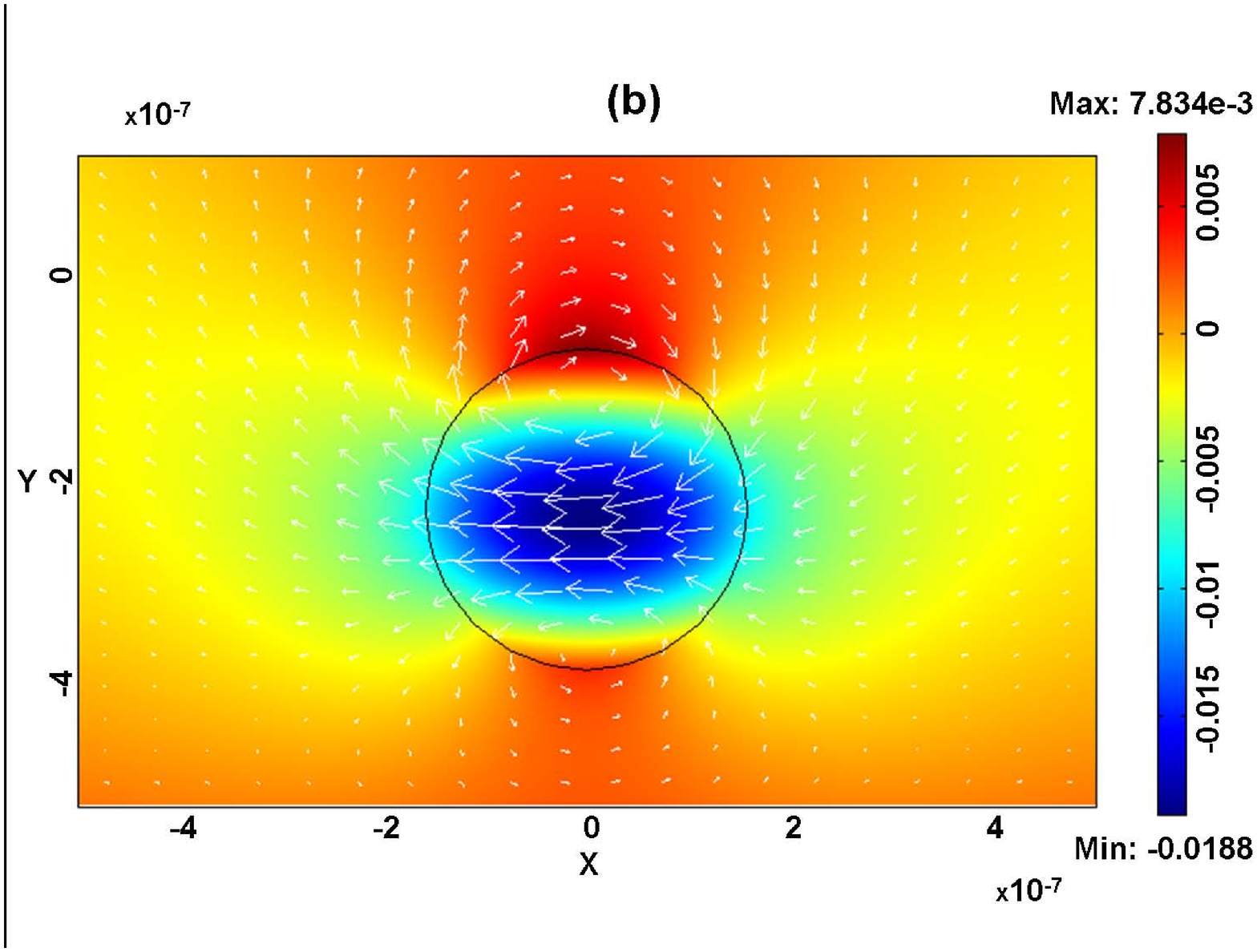}
\end{minipage}
\begin{minipage}{.98\linewidth}
\centering
\includegraphics[width=6cm]{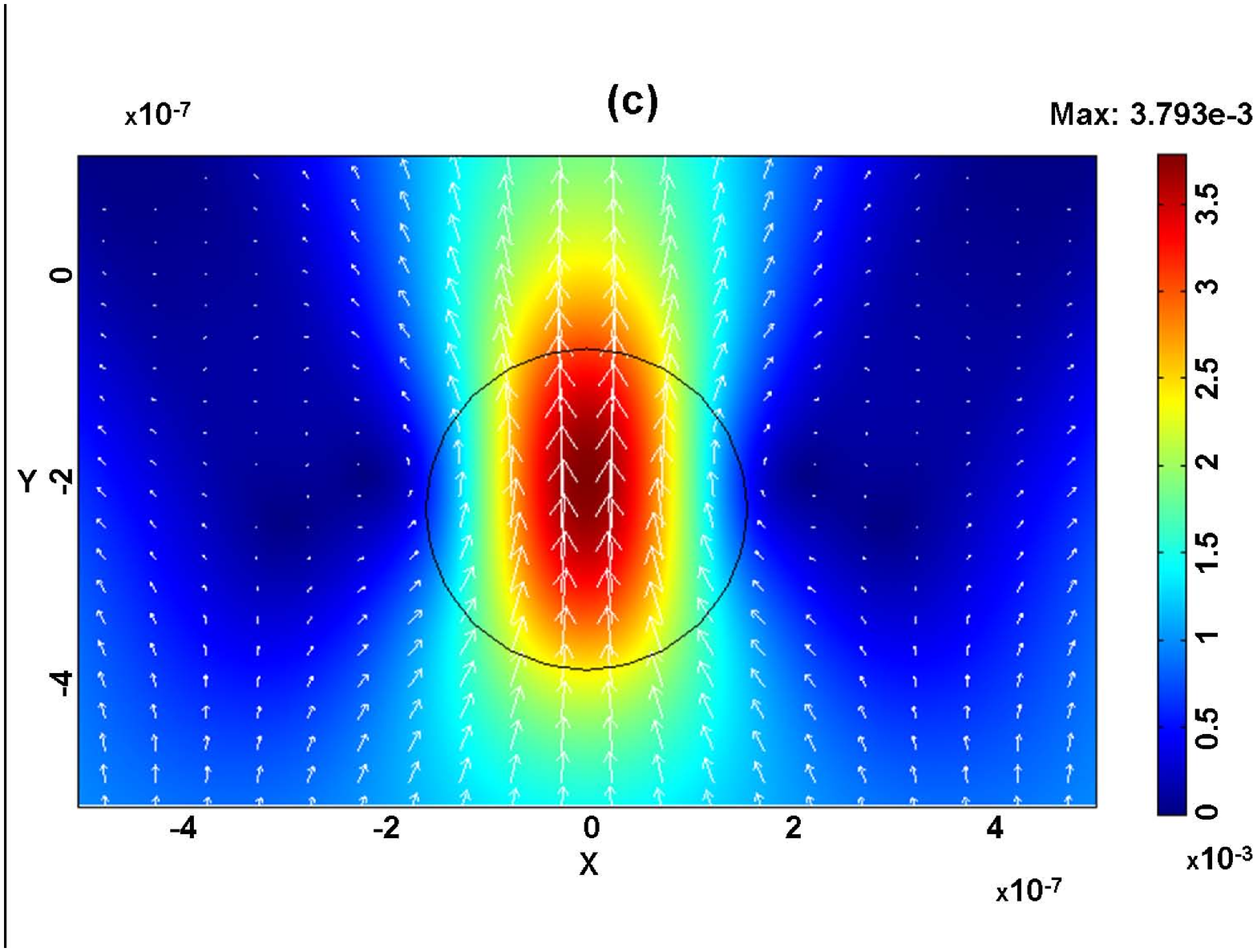}
\end{minipage}
\caption{ (a) Electric field $E_z({\bf R})$ in a cylinder of Si with
a dielectric permittivity $\epsilon= 12$ and radius $r= 158nm$. (b)
Magnetic field ${\bf H(R)}$ (arrows) and its X - component (colors).
(c) Averaged energy flow ${\bf <S(R)>}$ (arrows) and its norm
(colors). An s - polarized Gaussian beam of amplitude $A= 1V/m$ and
standard deviation $\sigma= 2792nm$ at $\lambda= 1.55\mu m$ is
launched upwards (i. e. with ${\bf R_i}$ along the OY direction),
from below the cylinder, exciting its $WGM: TM_{1,1}$.}
\end{figure}

On the other hand, Figs. 2(a) - 2(c) show the response of one single
Si cylinder to infrared light. The characteristics quoted above for
the BST ceramic cylinder are again reproduced for this Si rod, as
expected from the analysis of their similar electric and magnetic
resonances \cite{Vynck2009}.

\subsection{Transmission characteristics of a slab with ordered rod distributions in the microwave regime}
\begin{figure}[htbp]
\begin{minipage}{.49\linewidth}
\centering
\includegraphics[width=7cm]{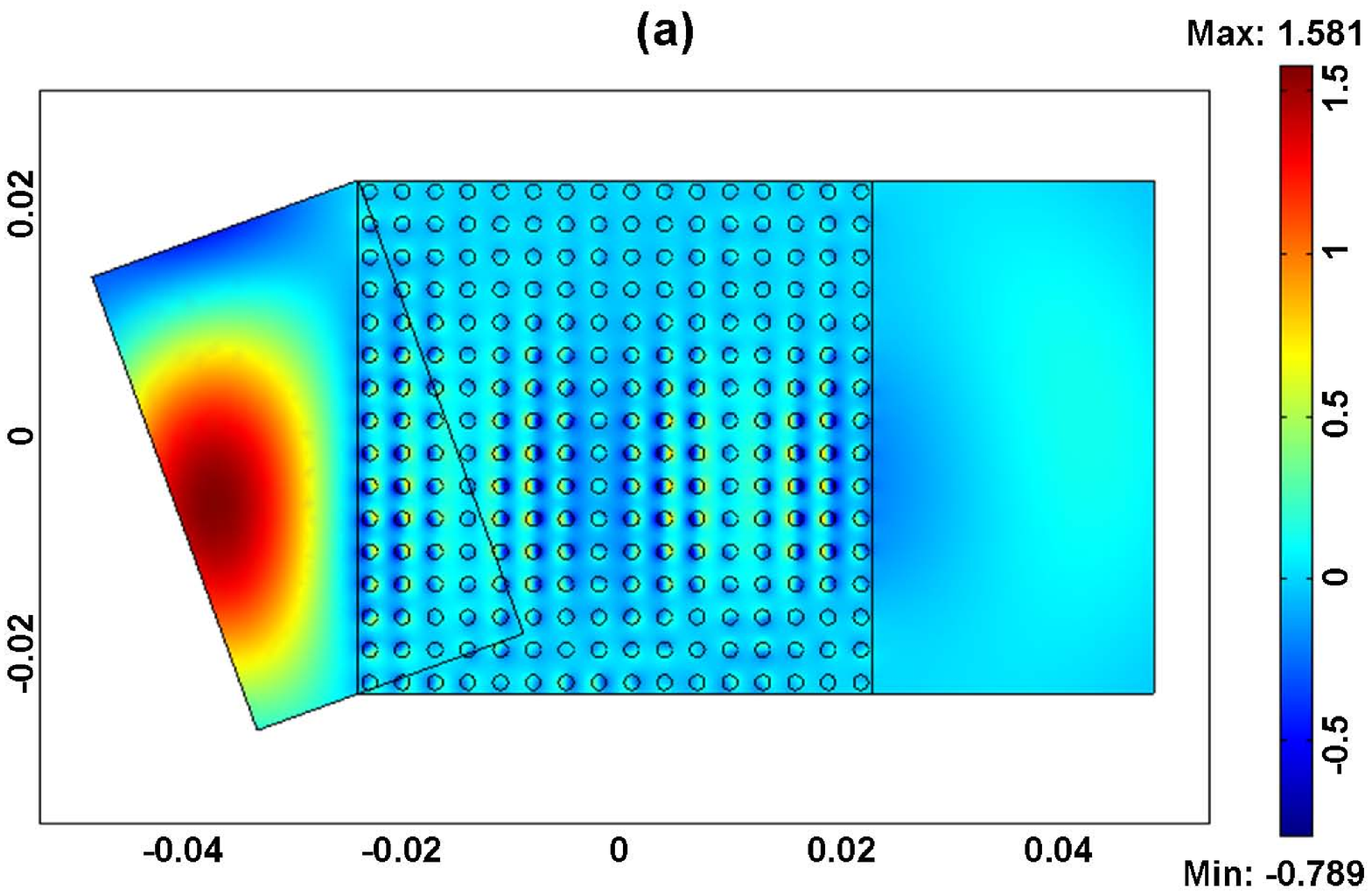}
\end{minipage}
\begin{minipage}{.49\linewidth}
\centering
\includegraphics[width=7cm]{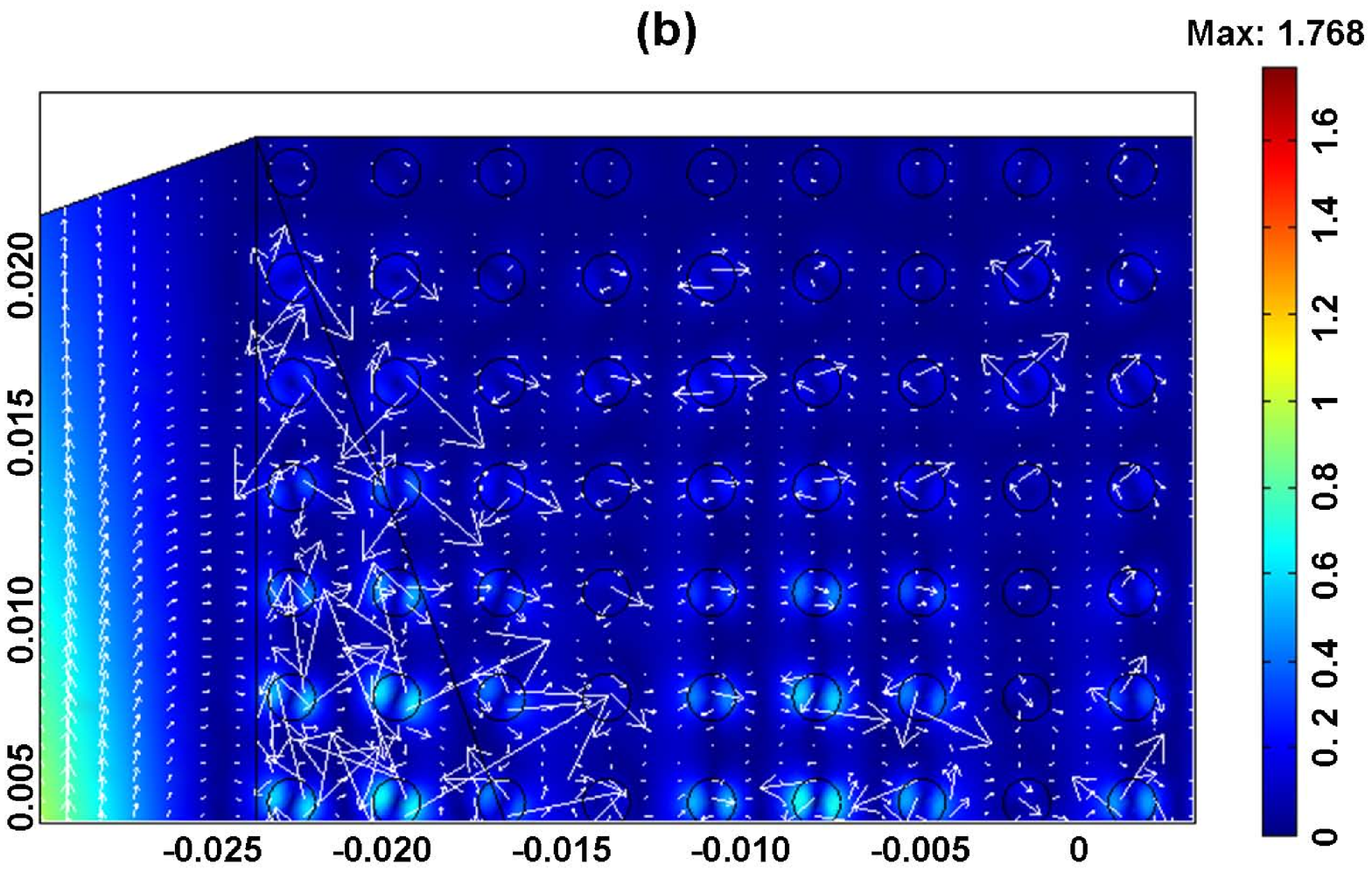}
\end{minipage}
\begin{minipage}{.49\linewidth}
\centering
\includegraphics[width=7cm]{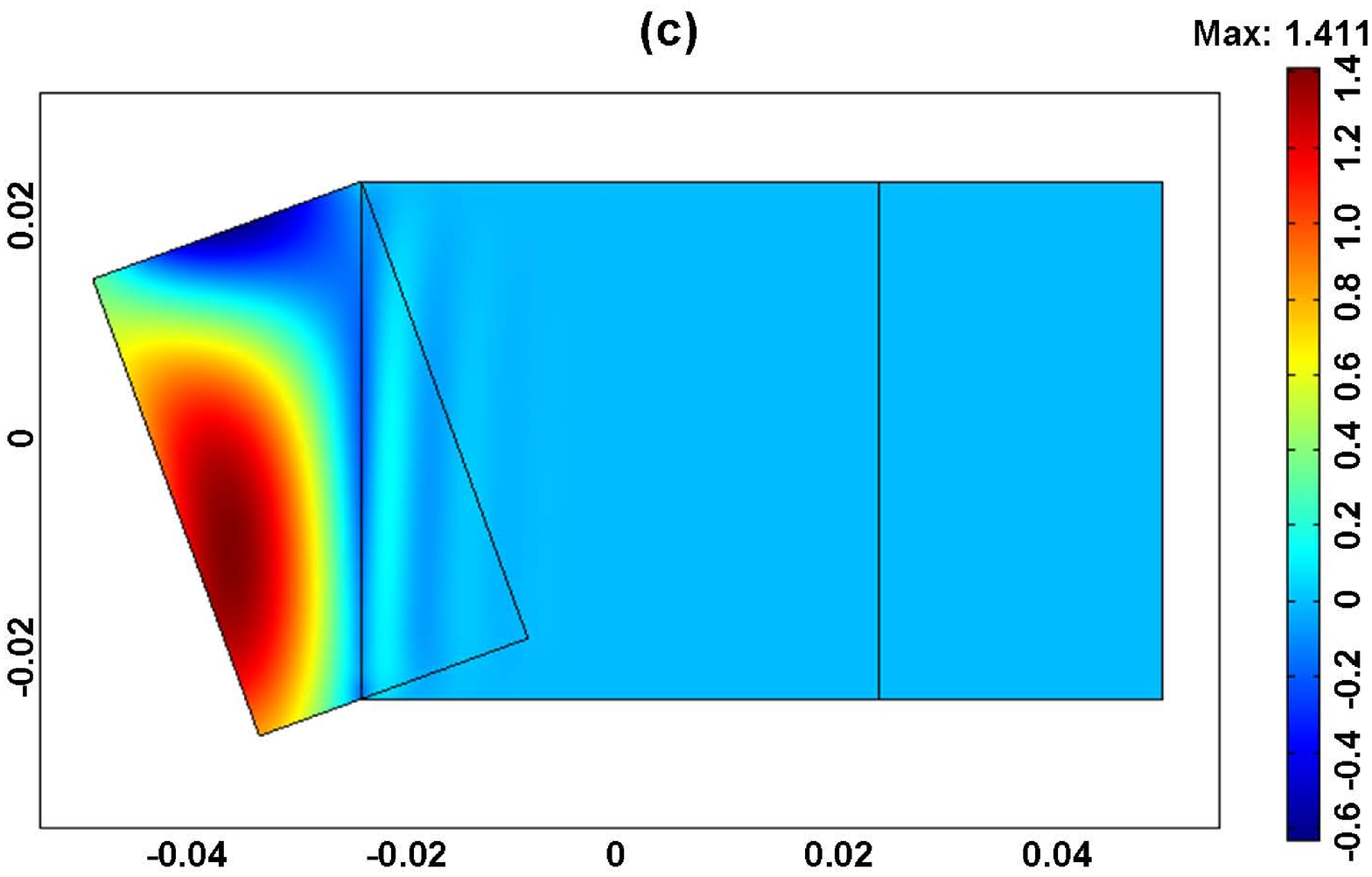}
\end{minipage}
\begin{minipage}{.49\linewidth}
\centering
\includegraphics[width=7cm]{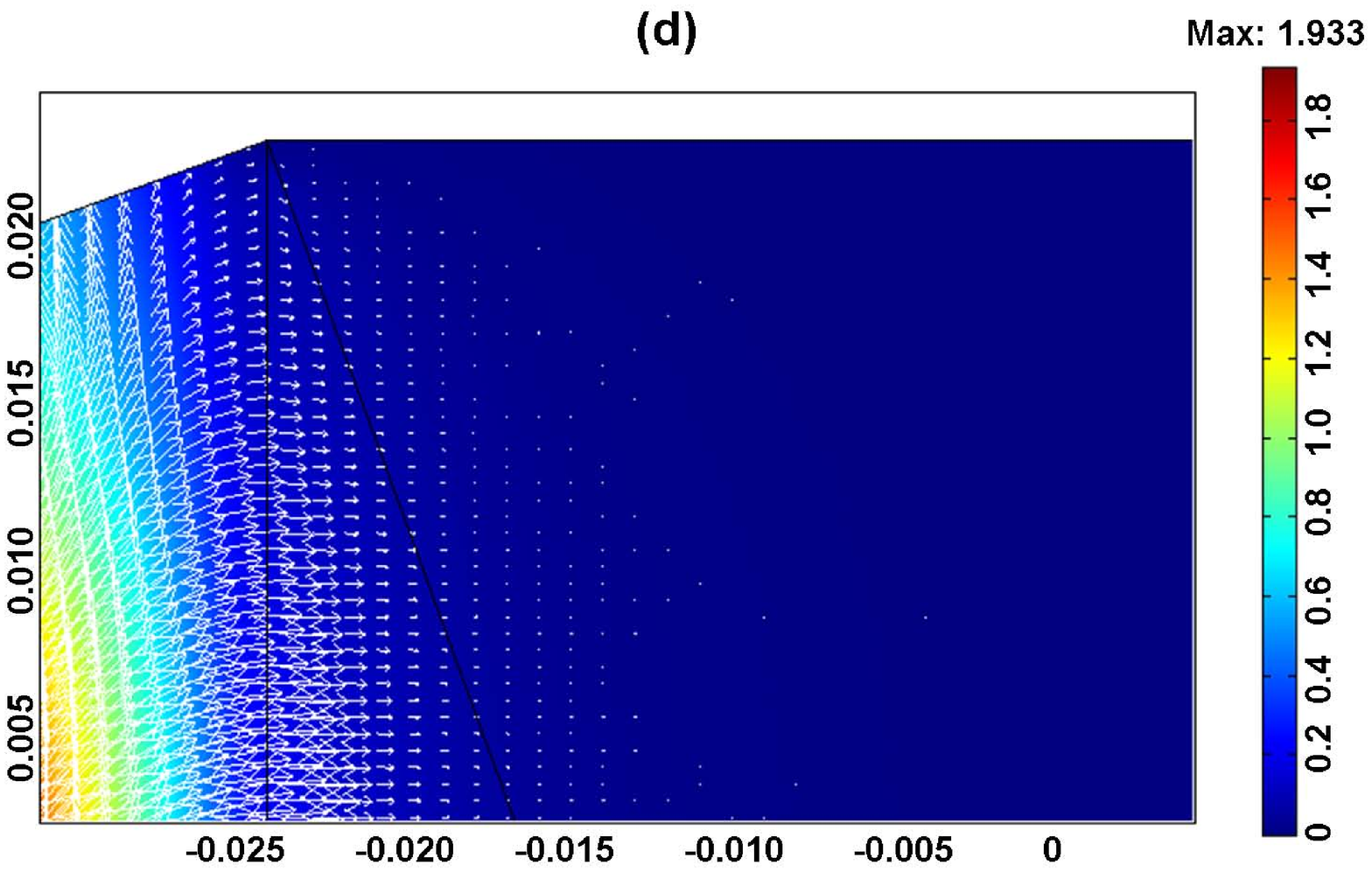}
\end{minipage}
\caption{ (a) Electric field $E_z({\bf R})$ spatial distribution of
a microwave propagating through a slab of an ordered array of BST
rods as the one of Figs. 1(a) - (c) ($\epsilon= 600$, $r= 0.68mm$).
The lattice constant is $a= 3mm$. An s - polarized Gaussian beam of
amplitude $A= 1V/m$ and width $\sigma= 4a$ is launched on the PC
from the left at wavelength $\lambda= 41.638mm$ ($\nu= 7.2GHz$) and
at an incidence angle $\theta= 20^\circ$ with the X - axis. (b)
Electric field norm $|E_z({\bf R})|$ (colors) and ${\bf <S(R)>}$
(arrows) in a detail of the block upper left corner. (c) Same as in
(a) in a slab occupied by a homogeneous medium whose effective
parameters obtained from the EMT are $\epsilon_{eff}= -36 + i14.7$,
$\mu_{eff}= -0.9 + i0.11$. (d) Electric field norm $|E_z({\bf R})|$
(colors) and ${\bf <S(R)>}$ (arrows) in a detail of the block upper
left corner.}
\end{figure}

We now consider a thick slab made of a periodic array of the BST
rods in air whose response was studied in Figs. 1(a) - (c), (see
Figs. 3(a) and 3(b)). An s - polarized light beam is launched on
this system at angle $\theta= 20^\circ$. The map of transmitted
$E_z({\bf R})$ field is shown in Fig. 3(a). A transmission angle
$\theta\approx 0^{\circ}$ is observed within the sample. A resonant
field distribution appears within the rods, being somewhat similar
to that of Fig. 1(a), taking into account the frequency shifts due
to the presence of neighbor rods. On the other hand, Fig. 3(b) shows
both the time - averaged electric energy $|E_z({\bf R})|$ and energy
flow ${\bf <S(R)>}$ in a detail of the block of rods. The latter
showing the Bragg directions of propagation inside the crystal.
Nevertheless, in the experiment of \cite{Peng2007} (cf. Figs. 3 of
Ref. \cite{Peng2007}) on a prism of an identical array, a refraction
angle of about $20^\circ$ was obtained instead, which would
correspond to $n_{eff} \simeq -1.08$. A 75\% of transmitted energy
is lost in the rod slab of Figs. 3(a) and 3(b).

\begin{figure}[htbp]
\centering
\includegraphics[width=7cm]{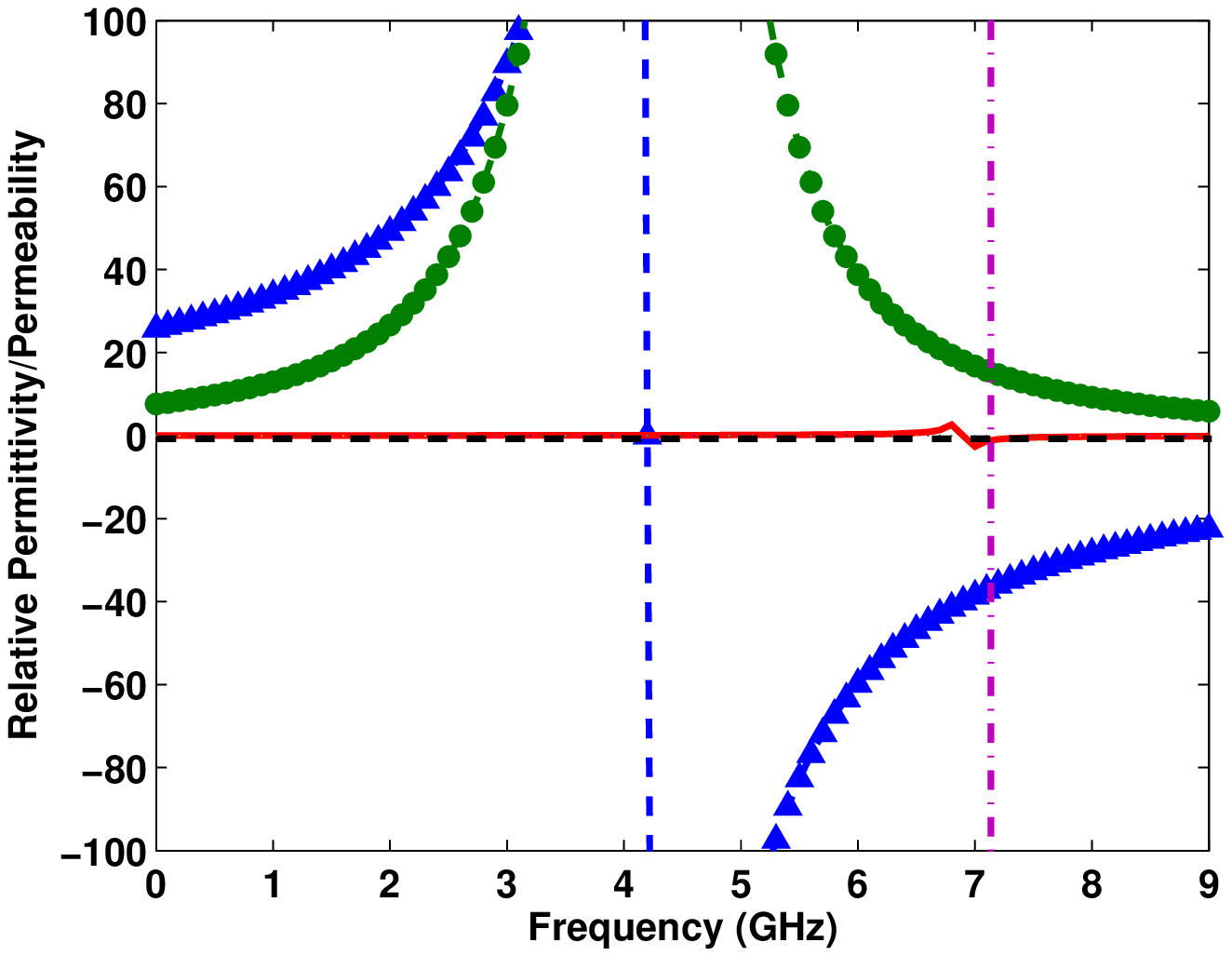}
\caption{Estimation of both the complex relative permittivity
$\epsilon_{eff}$ and permeability $\mu_{eff}$ for the BST rod array
in Fig. 3(a) obtained the effective medium theory.
$\epsilon^R_{eff}$ (broken line with triangles. Blue online),
$\epsilon^I_{eff}$ (broken line with dots. Green online),
$\mu^R_{eff}$ (full line. Red online) and $\mu^I_{eff}$ (broken
line. Black online). The broken with dots vertical line (violet
online) at frequency $\nu= 7.2GHz$ indicates both $\epsilon_{eff}$
and $\mu_{eff}$ values for the medium of Figs. 3(c) and 3(d):
$\epsilon_{eff}= -36$, $\epsilon^I_{eff}= 14.7$, $\mu^R_{eff}= -0.9$
and $\mu^I_{eff}= 0.11$. The full vertical (blue online) line
belongs to the jump of $\epsilon^R_{eff}$.}
\end{figure}

However, homogenization from an effective medium theory employed
like in \cite{Peng2007, Nieto2011, Lewin1948, Holloway2003} leads us
to estimate for this composite the effective homogeneous medium
constitutive parameters $\epsilon^R_{eff}= -36$, $\epsilon^I_{eff}=
14.7$; $\mu^R_{eff}= -0.9$, $\mu^I_{eff}= 0.11$ (see Fig. 4). Such a
uniform medium produces an angle of refraction which is very small,
similar to that of Fig. 3(a) but does not reproduce the inner
structure of the wavefield. This is shown in Figs. 3(c) and 3(d)
which display $E_z({\bf R})$ and a detail of $|E_z({\bf R})|$ and
${\bf <S(R)>}$ for such an effective medium.

Now a 95\% of losses appears in the transmitted energy emerging from
the homogeneous medium of Figs. 3(c) and 3(d). These results
manifest the very different observed refraction depending on the
sample on use: whether it is the composite or we deal with its EMT
homogenization. In fact, although not shown here, we remark that we
have reproduced the numerical results of \cite{Peng2007} for such an
array if the sample is a prism which, as said above, yields a
refraction completely different to that of Figs. 3(a) and 3(b).

Hence, we infer the questionable matching of an EMT with the
observations in such a regular array of rods. This is not surprising
in the light of the results of \cite{Zhang2011} and
\cite{Zhang2008}, since the sample in these cases is not much larger
than the inner structure, at difference with light transmission
experiments in usual light refractive elements of nanoscopic
molecular inner structure sizes.

%\newpage

\subsection{Microwave transmission in a slab with a random distribution of cylinders}

\begin{figure}[htbp]
\begin{minipage}{.49\linewidth}
\centering
\includegraphics[width=7cm]{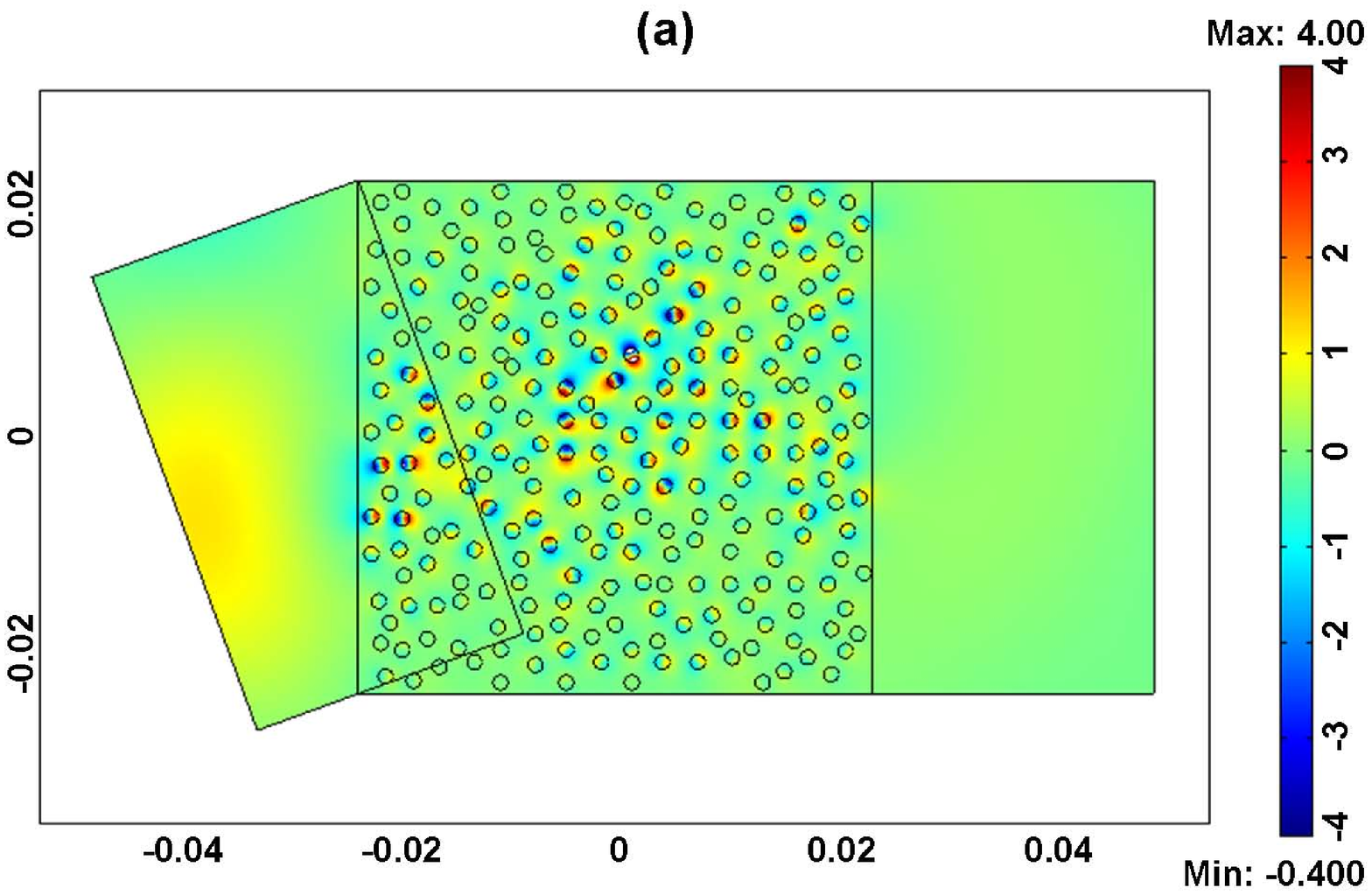}
\end{minipage}
\begin{minipage}{.49\linewidth}
\centering
\includegraphics[width=7cm]{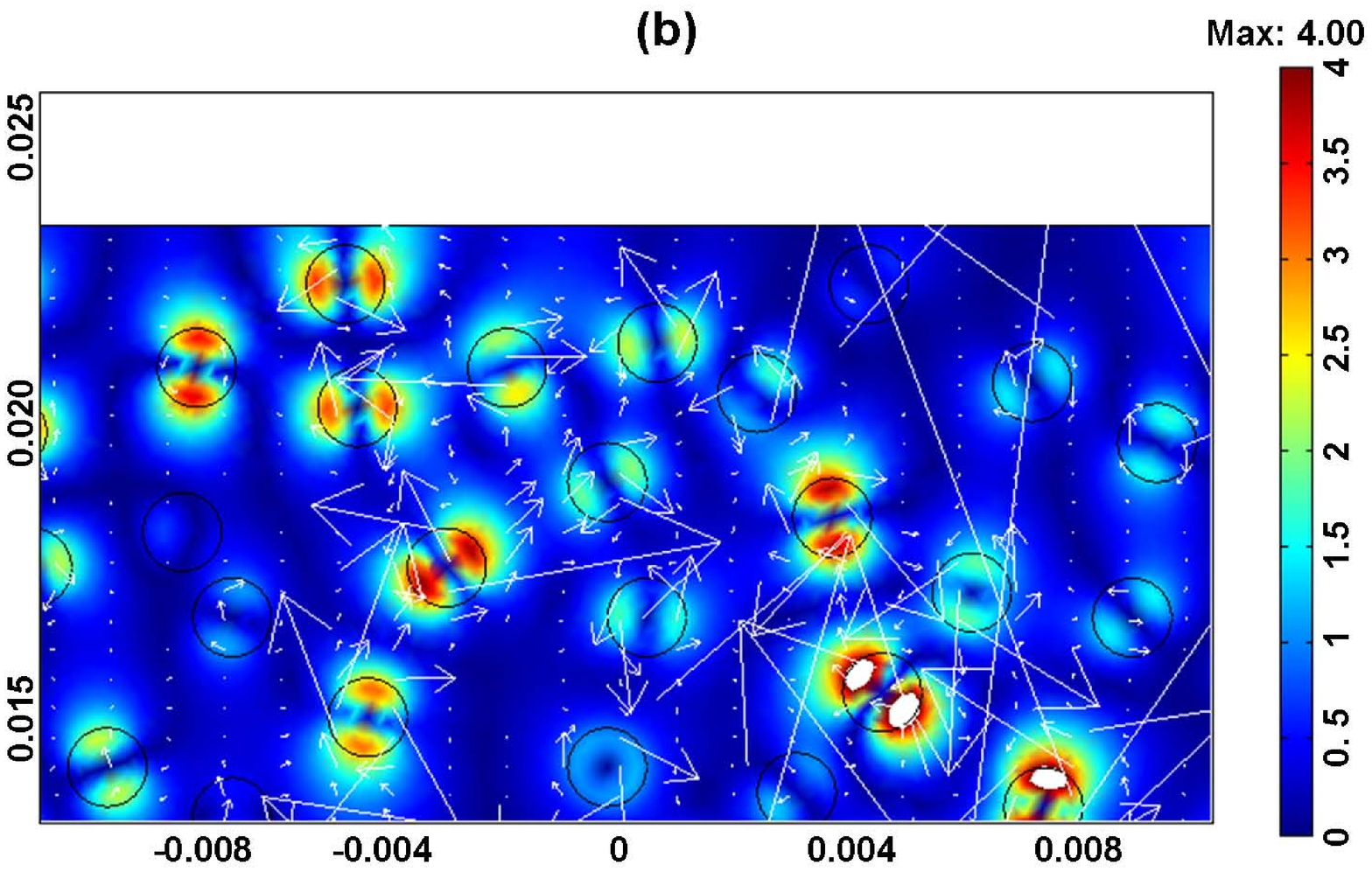}
\end{minipage}
\caption{ (a) Electric field $E_z({\bf R})$ spatial distribution in
a random configuration of cylinders like those of Figs. 1(a) - (c)
keeping the same filling fraction and the same conditions of
illumination as in the ordered array of Figs. 3. (b) Electric field
norm $|E_z({\bf R})|$ (colors) and ${\bf <S(R)>}$ (arrows) in a
detail of the block central region which includes its upper
transparent side.}
\end{figure}

{\noindent Different realizations, obtained by randomizing the}
array of Figs. 3(a) and 3(b) are now addressed. We next study a
disordered distribution of BST rods as those employed in Section
~\ref{ORS_MW}. Now, there is 85\% of transmission losses through the
slab of these disordered rods due to extinction produced by
scattering. Figs. 5(a) and 5(b) show the field $E_z({\bf R})$ and a
detail of both $|E({\bf R})|$ and ${\bf <S(R)>}$, respectively. A
large amount of scattered light is lost both above and below the
slab upper and lower low reflection boundaries. Again some few rods
exhibit the excitation of Mie resonances on illumination as in Fig.
1(a). Figures 5(a) and 5(b) yield no clue of a forward transmission
direction into the air at the exit of the sample. This result is
obtained with both a beam and a plane wave. Although not shown here,
we should state that we observed that averaging the field $E_z({\bf
R})$ over many realizations of the random distribution of rods does
not yield a forwardly transmitted beam, characterized by
$<E_{z}({\bf R})>$ \cite{Ishimaru, Nieto1997, Nieto2000, Nieto2001},
distinguishable from the scattered light.

\subsection{Ordered and random distributions of Silicon rods in the mid - infrared regime}

\begin{figure}[htbp]
\begin{minipage}{.49\linewidth}
\centering
\includegraphics[width=7cm]{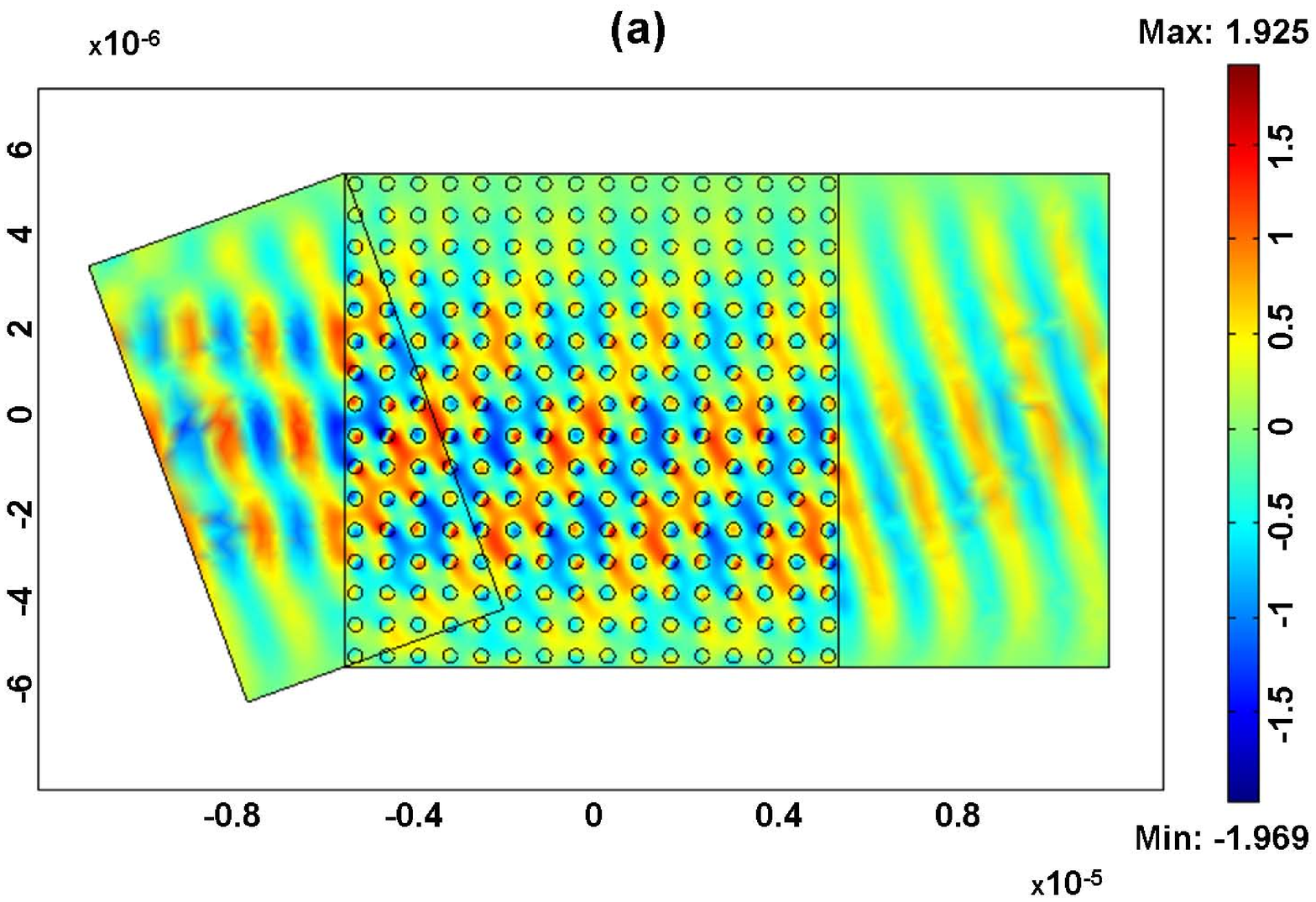}
\end{minipage}
\begin{minipage}{.49\linewidth}
\centering
\includegraphics[width=7cm]{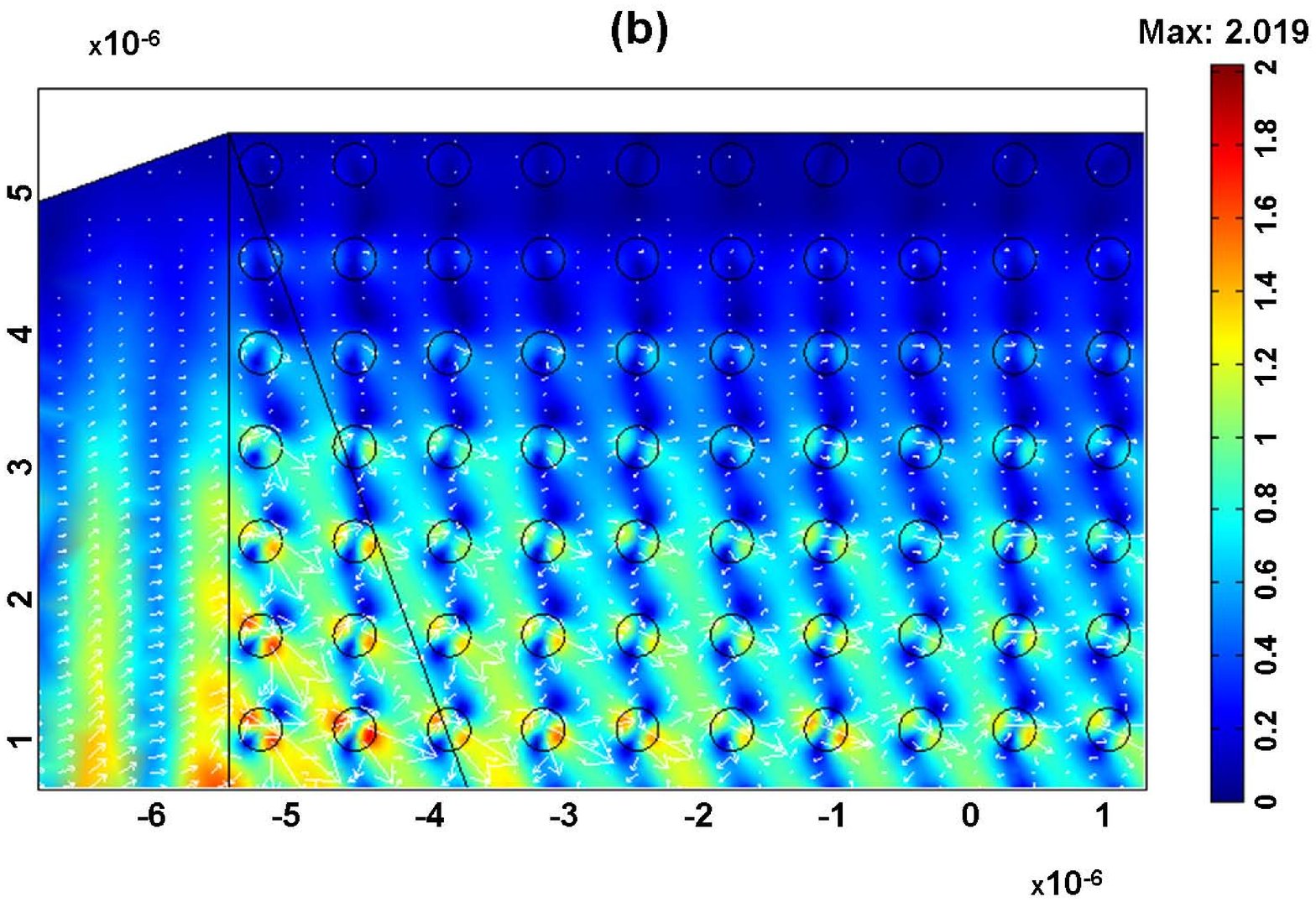}
\end{minipage}
\begin{minipage}{.49\linewidth}
\centering
\includegraphics[width=7cm]{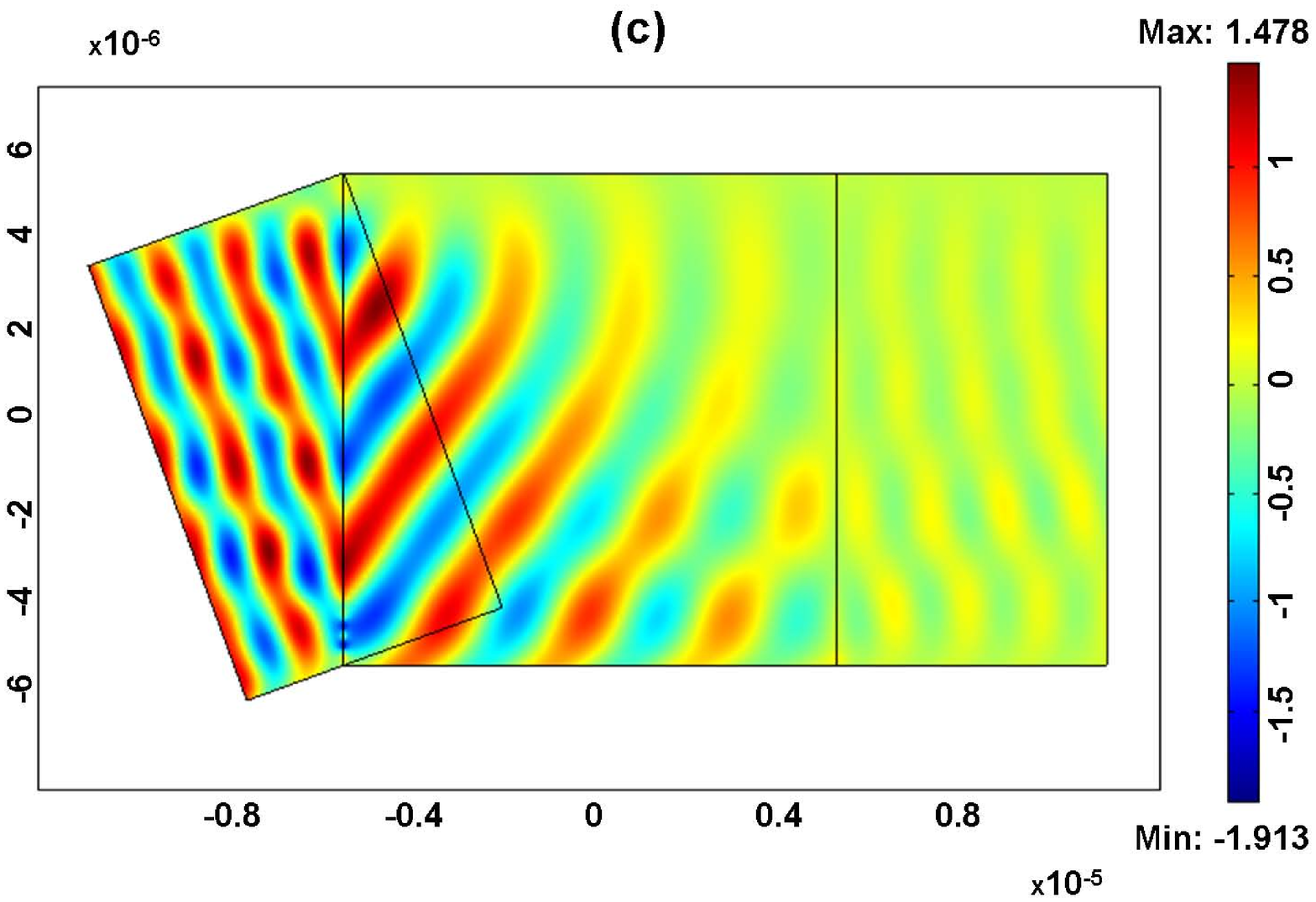}
\end{minipage}
\begin{minipage}{.49\linewidth}
\centering
\includegraphics[width=7cm]{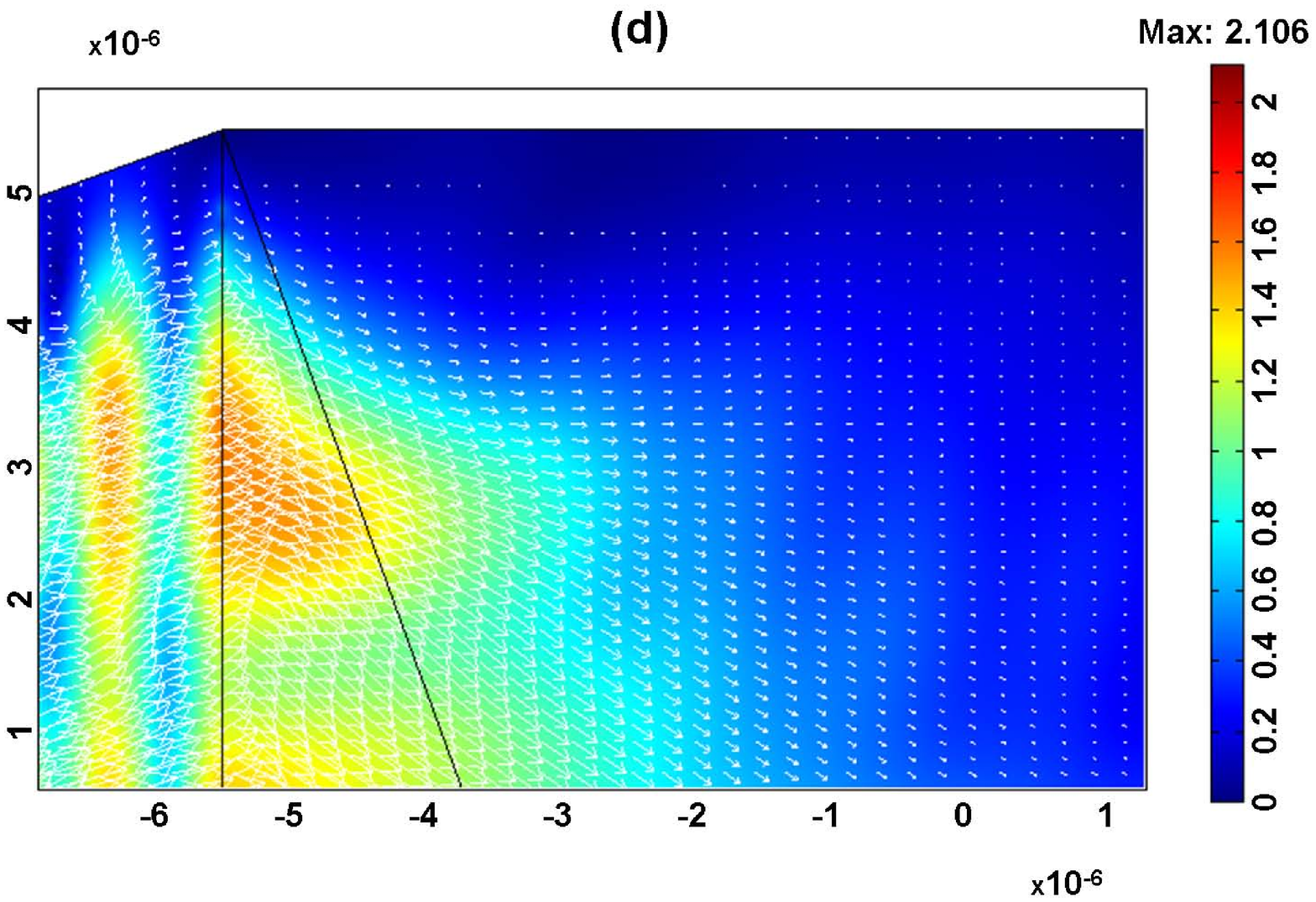}
\end{minipage}
\caption{ (a) Electric field $E_z({\bf R})$ propagating in a thick
slab occupied by an ordered array of Si rods ($\epsilon= 12$ and $r=
158nm$; the lattice constant is $a= 698nm$). An s - polarized
Gaussian beam of amplitude $A= 1V/m$, $\sigma= 4a$ and wavelength
$\lambda= 1.55\mu m$ is launched on the slab from the left at an
incidence angle $\theta= 20^\circ$. (b) Electric field norm
$|E_z({\bf R})|$ (colors) and ${\bf <S(R)>}$ (arrows) in a detail of
the block upper left corner. (c) Electric field $E_z({\bf R})$ in a
uniform medium optically equivalent to that of (a) and (b), whose
electric permittivity is $\epsilon_{eff}= -0.36 + i0.25/8$. The real
part of such a value is estimated by Snell's law, ($\theta_{i}=
21.54^\circ$, $\theta_{t}= 38.07^\circ$). The imaginary part has
been estimated from the transmittivity of Figs. 6(a) and 6(b). (d)
Electric field norm $|E_z({\bf R})|$ (colors) and ${\bf <S(R)>}$
(arrows) in a detail of the block upper left corner.}
\end{figure}

We next address a thick slab of ordered Si cylinders like that
studied in Figs. 2. This arrangement is an extension to mid - IR
(\cite{ Vynck2009}) of the previous calculations at microwaves
discussed in Sections ~\ref{ORS_MW} and ~\ref{RRS_MW}. As shown in
Figs. 6(a) and 6(b) there appears negative refraction (i. e. a
backward wave) in the block and no appreciable transmission losses.
The positive and negative peak values of the wavefronts transmitted
inside the array coincide with those cylinders that appear
resonantly illuminated thus exhibiting the Mie $T_{1, 1}$ resonance
like in Fig. 2(a). Mainly, there are ${\bf <S(R)>}$ arrows pointing
in the direction normal to these wavefronts. Light is transmitted
into the air side on the right through this slab with the same angle
as that of incidence. This confirms the analysis of \cite{Vynck2009}
for this crystal.

On the other hand, Figs. 6(c) and 6(d) display the transmission of
the same incident wave through a thick slab occupied by a uniform
medium with a refractive index: $n= -0.36 + i0.25/8$ which resembles
the propagation through the ordered array shown in Figs. 6(a) and
6(b). The real part $n^R= -0.36$ has been estimated from Snell's law
applied to Fig. 6(a), whereas the imaginary part $n^I= 0.25/8$ was
fitted to a transmittance of this homogeneous medium being
approximated to that of Figs. 6(a) and 6(b). Notice that the
transmittivity of the homogeneous slab of Figs. 6(c) and 6(d) is
smaller than that of the ordered array of Figs. 6(a) and 6(b).
However $n^I$ cannot decrease much beyond 0.25/8, since otherwise
well known instabilities due to divergences at the right side of the
slab \cite{Heyman2001, Nieto2002, Nieto2004} appear. However, it
should be remarked that this value of $n$ does not coincide with
that of an EMT, which according to the values of $a/\lambda$ and
$r/\lambda$ for this array, as discussed in Section ~\ref{Intro},
are too large for a homogenization procedure to work with such a
structure. This is further discussed next by employing other arrays
of these cylinders with the same filling fraction and lattice
parameter; namely, a thick slab of disordered Si rods and a prism of
ordered, or disordered, cylinders.

\begin{figure}[htbp]
\begin{minipage}{.49\linewidth}
\centering
\includegraphics[width=7cm]{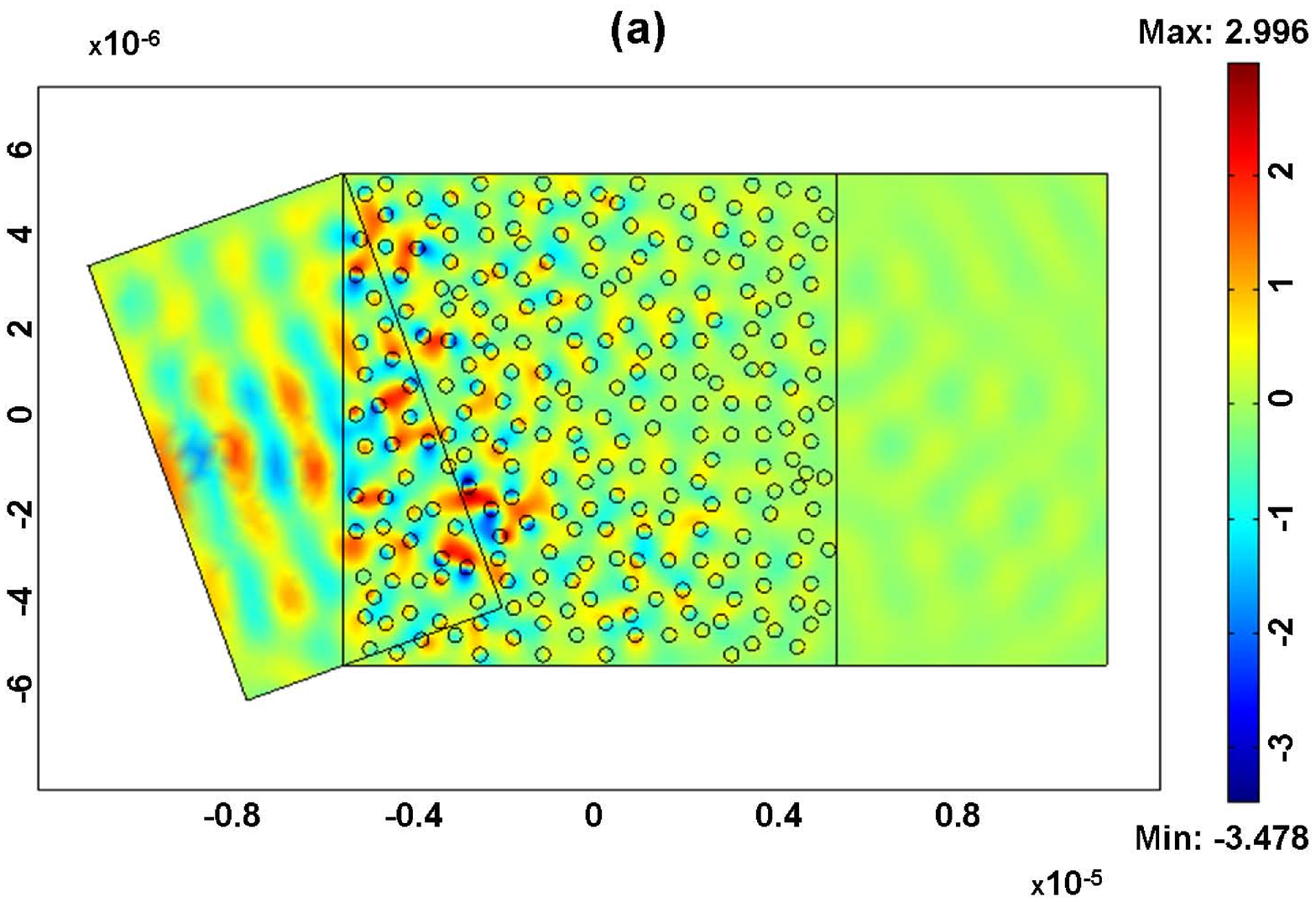}
\end{minipage}
\begin{minipage}{.49\linewidth}
\centering
\includegraphics[width=7cm]{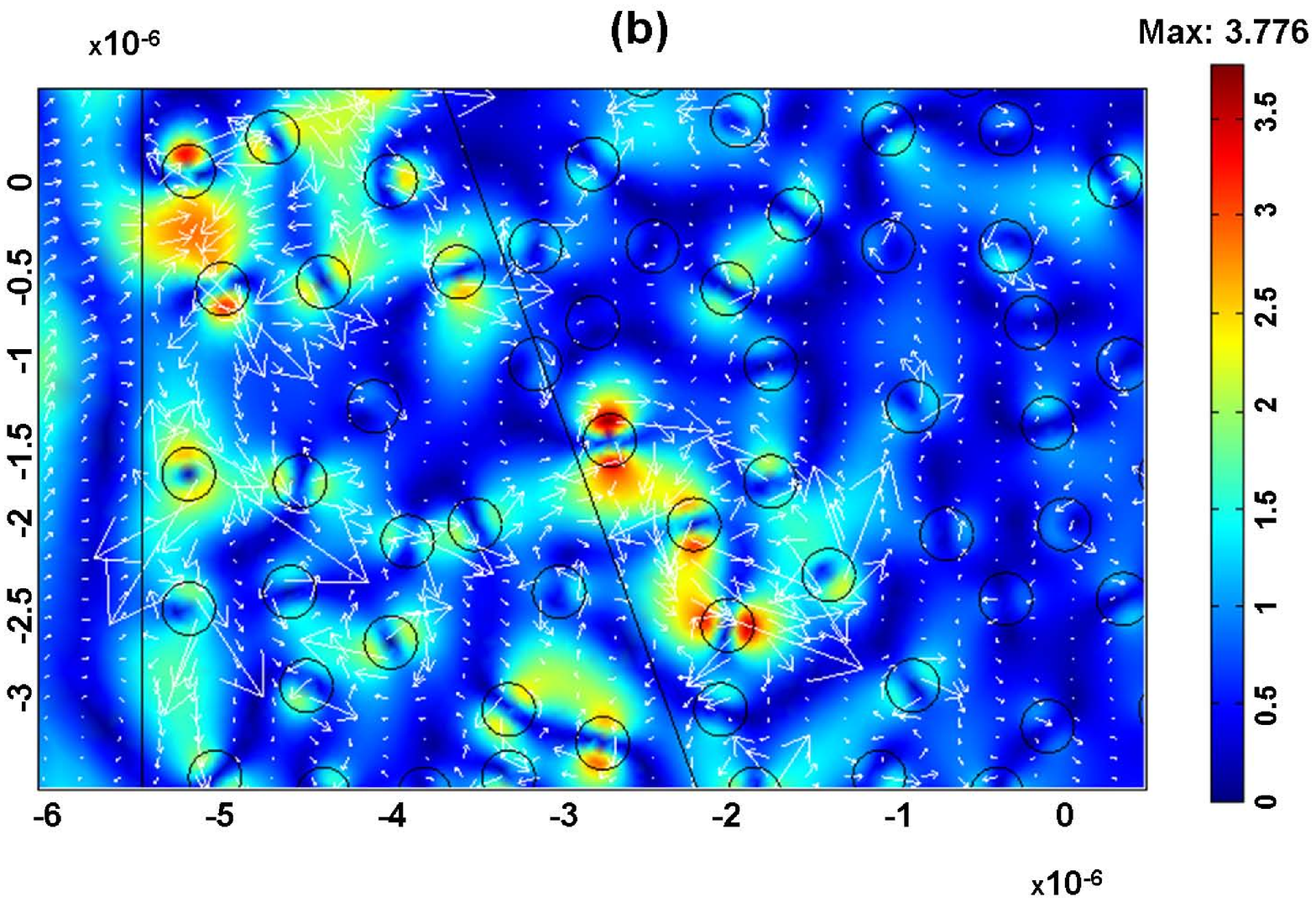}
\end{minipage}
\begin{minipage}{.98\linewidth}
\centering
\includegraphics[width=7cm]{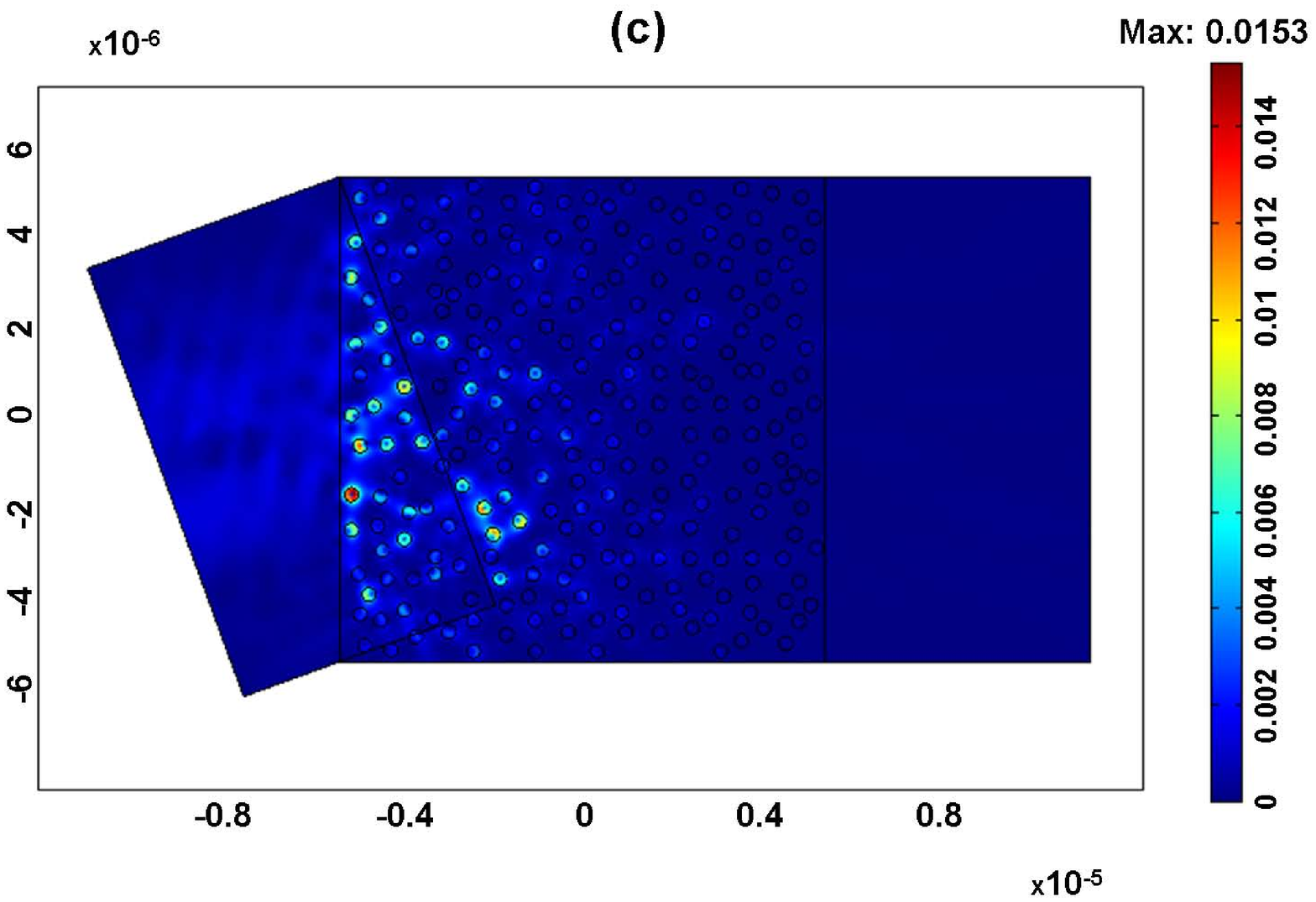}
\end{minipage}
\caption{ (a) Electric field $E_z({\bf R})$ when the configuration
of Si rods shown in Figs. 6(a) and 6(b) is disordered, keeping the
same filling fraction $f= 0.30$ and the same conditions of
illumination. (b) Electric field norm $|E_z({\bf R})|$ (colors) and
${\bf <S(R)>}$ (arrows) in an inset of the block left region which
includes its left interface. (c) Map of the averaged energy flow
norm ${\bf |<S(R)>|}$.}
\end{figure}

When these cylinders are disordered keeping $f\approx0.30$, as seen
in Figs. 7(a) - (c), there is no such negative refraction inside the
block as that shown in Figs. 6. Now there is a huge extinction of
energy due to high scattering by the particles. By averaging over
several realizations, there is no observed refractive transmission
of a forward component $<{\bf E_{z}}>$ into the air side. The block
of disordered rods now scatters, see Figs. 7(a) and 7(b), more than
90\% of the transmitted intensity both above and below its upper and
lower low reflection boundaries. Hence, only some few spheres are
illuminated showing their $TM_{1, 1}$ Mie resonances. Figure 7(c)
shows that ${\bf |<S(R)>|}$ is transmitted into the air region at
the right of the slab with less than a 1/10 of its incident wave
value.

\subsection{A prism of ordered or disordered Si distributions in the infrared}

\begin{figure}[htbp]
\begin{minipage}{.49\linewidth}
\centering
\includegraphics[width=7cm]{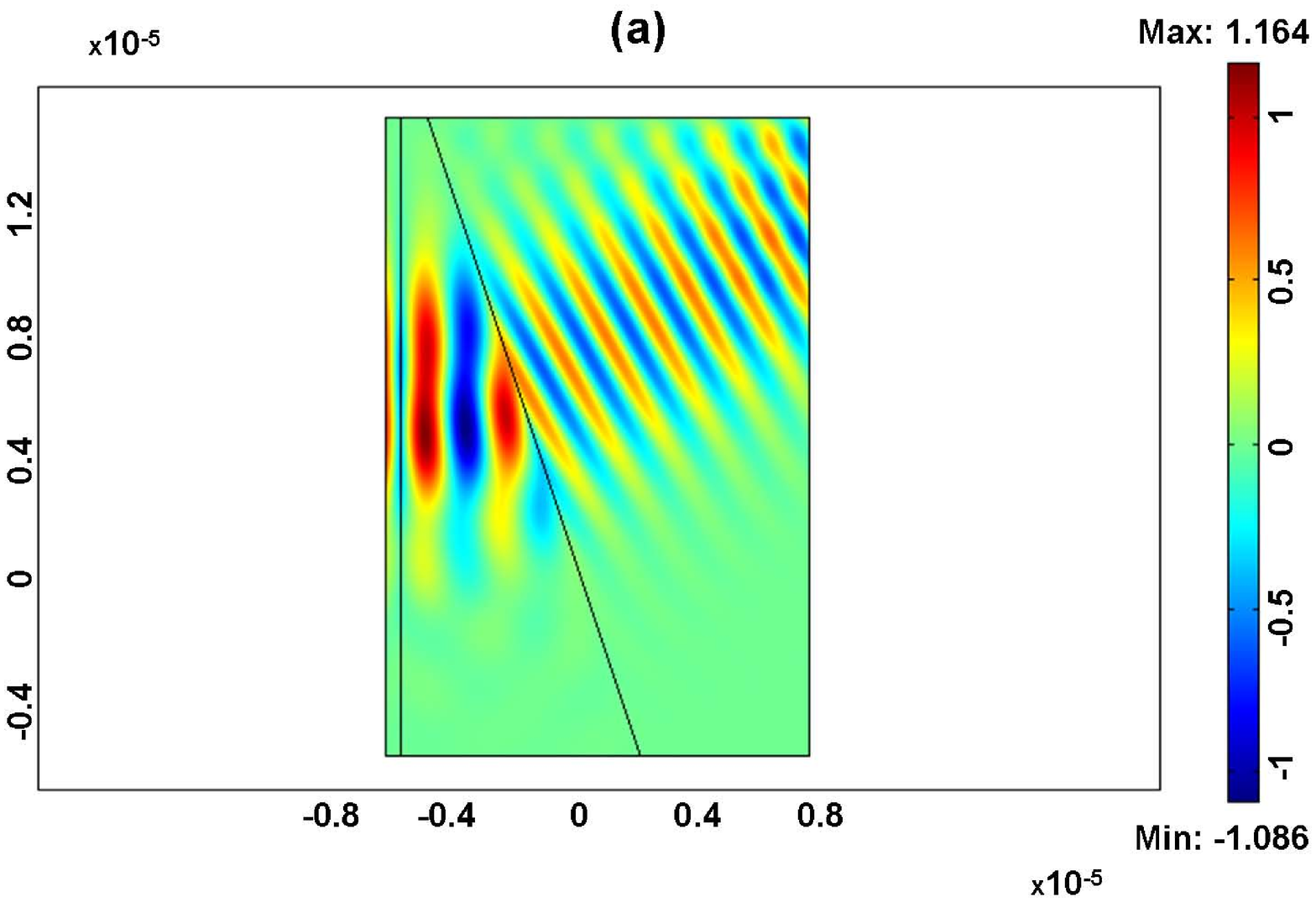}
\end{minipage}
\begin{minipage}{.49\linewidth}
\centering
\includegraphics[width=7cm]{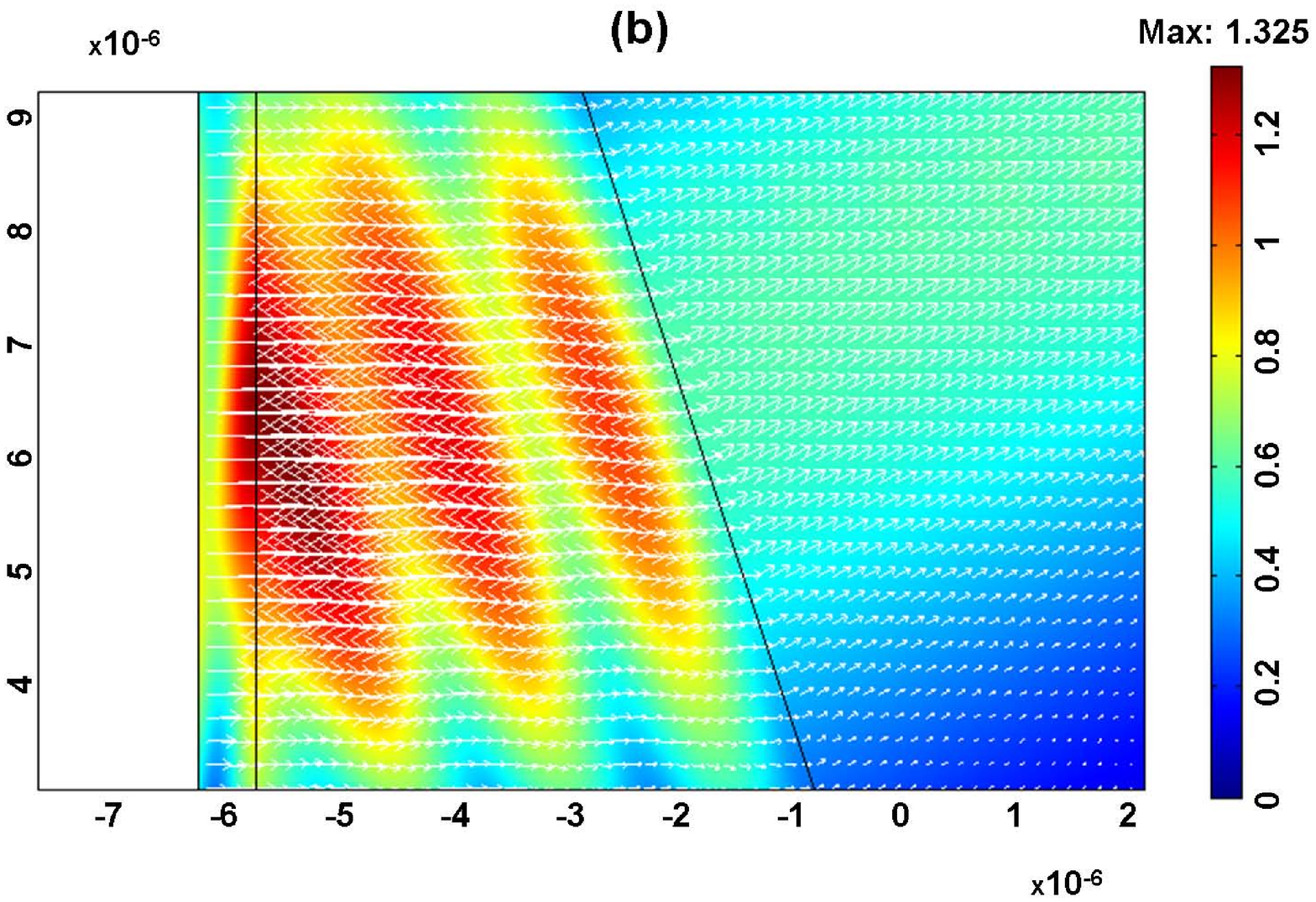}
\end{minipage}
\begin{minipage}{.98\linewidth}
\centering
\includegraphics[width=7cm]{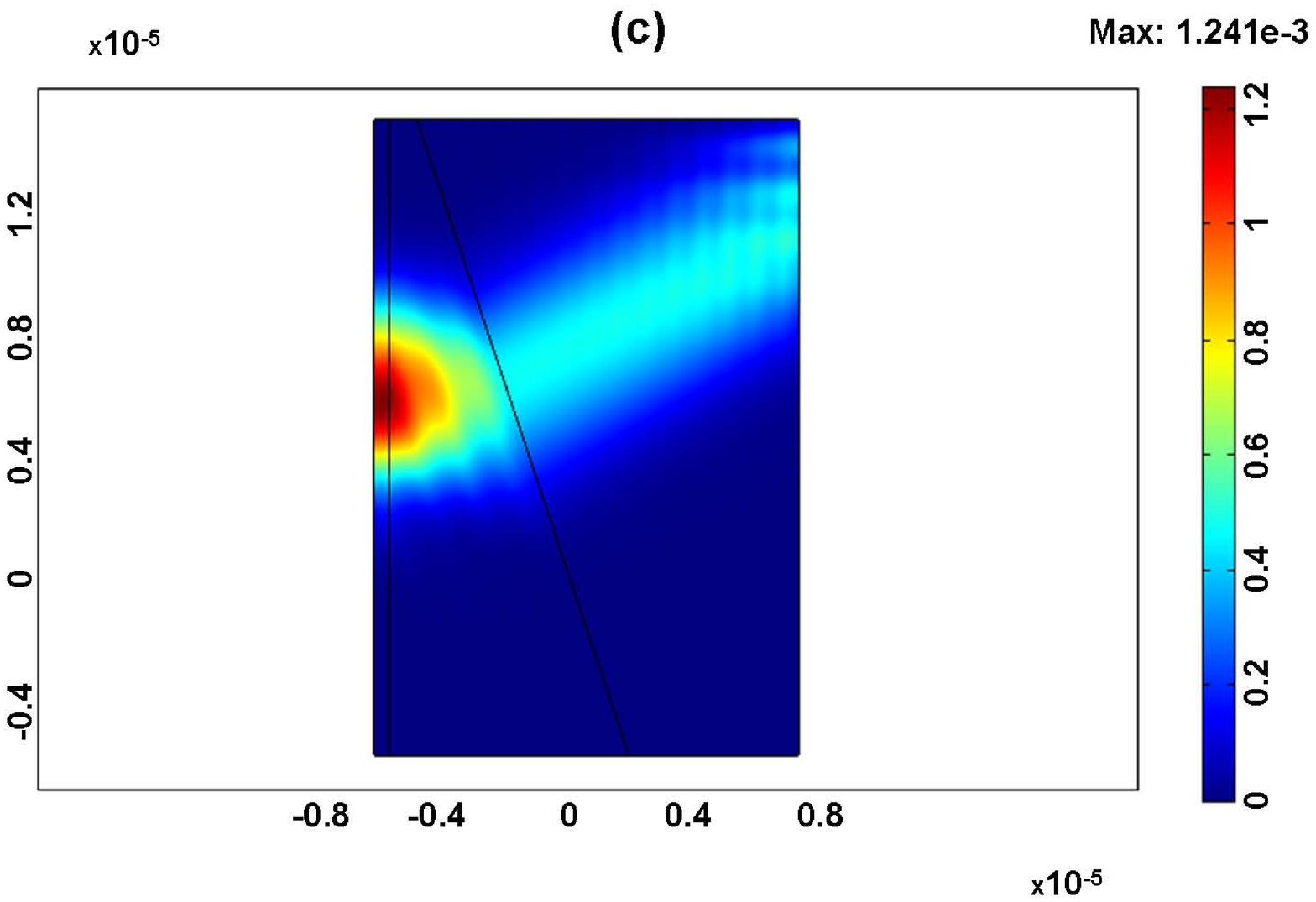}
\end{minipage}
\caption{ (a) Electric field $E_z({\bf R})$ in a prism (top angle
$\alpha= 18.4^\circ$) occupied by a uniform medium whose electric
permittivity is $\epsilon_{eff}= -0.36 + i0.25/8$ illuminated from
the left. This prism is optically as dense as the homogeneous thick
slab of Figs. 6(c) and 6(d). (b) Electric field norm $|E_z({\bf
R})|$ (colors) and ${\bf <S(R)>}$ (arrows) in a detail of the prism
upper region. (c) Map of the averaged energy flow norm ${\bf
|<S(R)>|}$. An s - polarized Gaussian beam of amplitude $A= 1V/m$,
$\sigma= 4\times 698nm$ and wavelength $\lambda= 1.55\mu m$ is
launched from the left on the prism at $\theta_i= 0^\circ$ with its
left side.}
\end{figure}

{\noindent Let us consider now a prism of a uniform medium of}
refractive index $n= -0.36 + i0.25/8$. Notice that this value of $n$
is the one of Figs. 6(c) and 6(d), derived from refraction at the
ordered array of Figs. 6(a) and 6(b). As shown in Figs. 8,
refraction into the air takes place with $\theta_{i}= 18.4^{\circ}$
and $\theta_{t}= 38.07^{\circ}$, the latter being the negative angle
of refraction at the prism larger side. This is shown in Figs. 8(a)
- (c) and serves us as a reference to study transmission through
both an ordered and disordered array of Si rods contained in a
sample with this prism geometry, which also was the one employed in
transmission observations at microwaves in \cite{Peng2007}.

\begin{figure}[htbp]
\begin{minipage}{.49\linewidth}
\centering
\includegraphics[width=7cm]{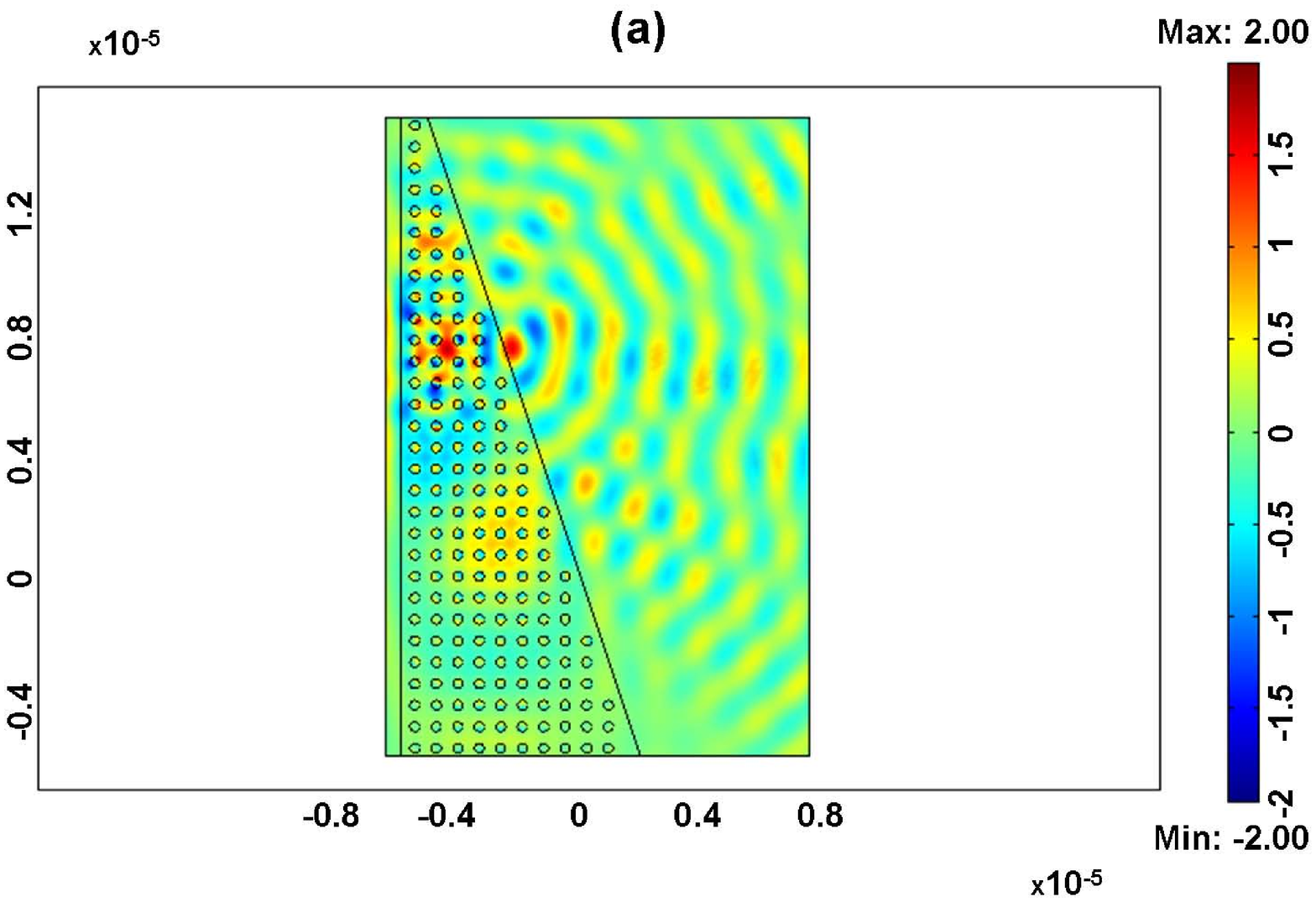}
\end{minipage}
\begin{minipage}{.49\linewidth}
\centering
\includegraphics[width=7cm]{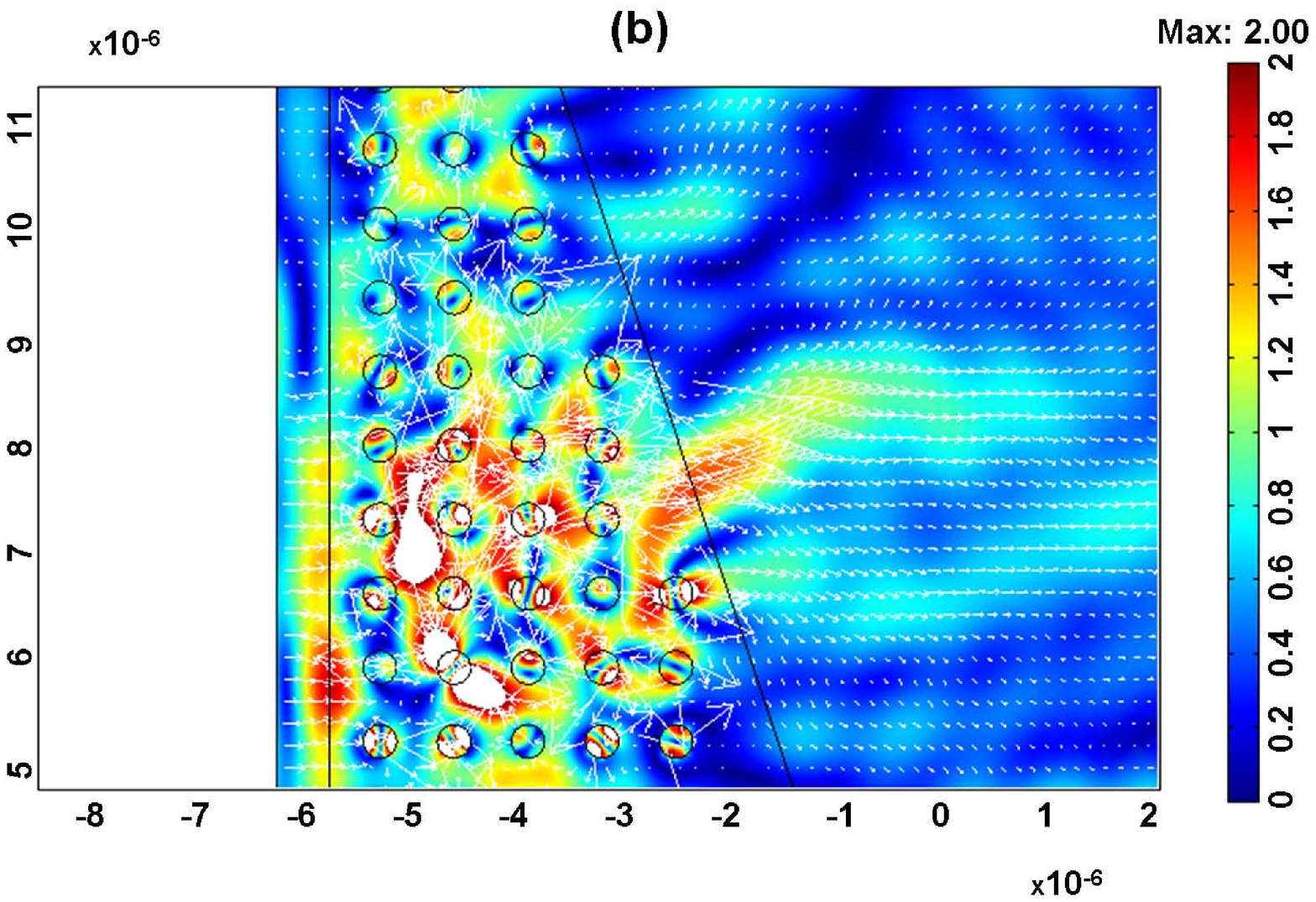}
\end{minipage}
\begin{minipage}{.98\linewidth}
\centering
\includegraphics[width=7cm]{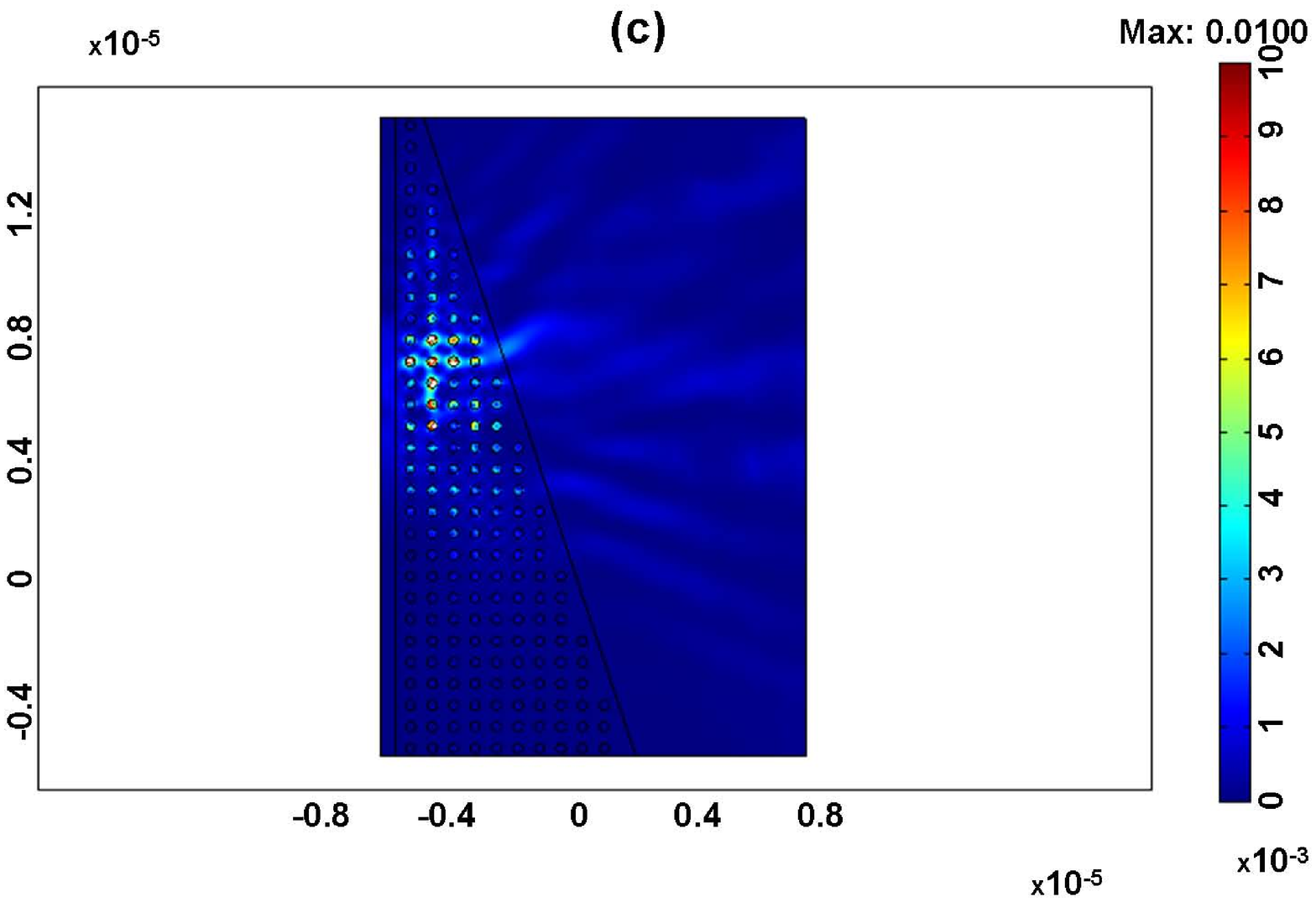}
\end{minipage}
\caption{ (a) Electric field $E_z({\bf R})$ in a prism with the same
shape as that of Figs. 8(a) - (c), at the same illumination
conditions, but now occupied by an ordered Si rod array as that of
Figs. 6(a) and 6(b). (b) Electric field norm $|E_z({\bf R})|$
(colors) and ${\bf <S(R)>}$ (arrows) in a detail of the prism upper
region. (c) Averaged energy flow norm ${\bf |<S(R)>|}$. The
illumination is the same as in Figs. 8(a) - 8(c).}
\end{figure}

Figures 9(a) - (c) show the wave propagation on illumination of the
prism filled with an ordered array of Si cylinders identical to that
of the thick slab of Figs. 6(a) and 6(b). As shown, there is now
absence of negative refraction at the larger side interface of this
prism. This contrasts with the observation in the sample with the
same shape filled with a uniform medium, as displayed in Figs. 8(a)
- (c) and also with the case of a thick slab filled with the same
array, as seen in Figs. 6(a) and 6(b). Hence, at difference with the
crystal of Si rods in the block, the same crystal in the prism does
not reproduce negative refraction, but rather a set of diffracted
orders into the air which are associated to the prism angle
$\alpha$, according to the conservation of the transversal
wavevectors at the larger interface. Also, the wave propagation
inside the prism, which should be like that appearing in the thick
slab of Fig. 6(a), but now at the same direction as the incident
wave, (since now this latter wave incides on the prism at $\theta_i=
0^\circ$), is observed in Figs. 9(a) and 9(b) to be quite different,
with a complicated structure due to the interference of the
different Bragg waves. This result, once again, points out the
dependence of wave propagation not only on the inner structure of
the composite, but also on its shape. In addition, these
observations demonstrate that the negative refraction found in the
slab of ordered Si rods, (cf. Fig. 6(a)) is not due to an effective
homogeneous medium effect, but it rather comes from a consequence of
the diffraction in the array of Si rods due to its lattice symmetry.
This answers the question posed in \cite{Zhang2009_2}.

\begin{figure}[htbp]
\begin{minipage}{.49\linewidth}
\centering
\includegraphics[width=7cm]{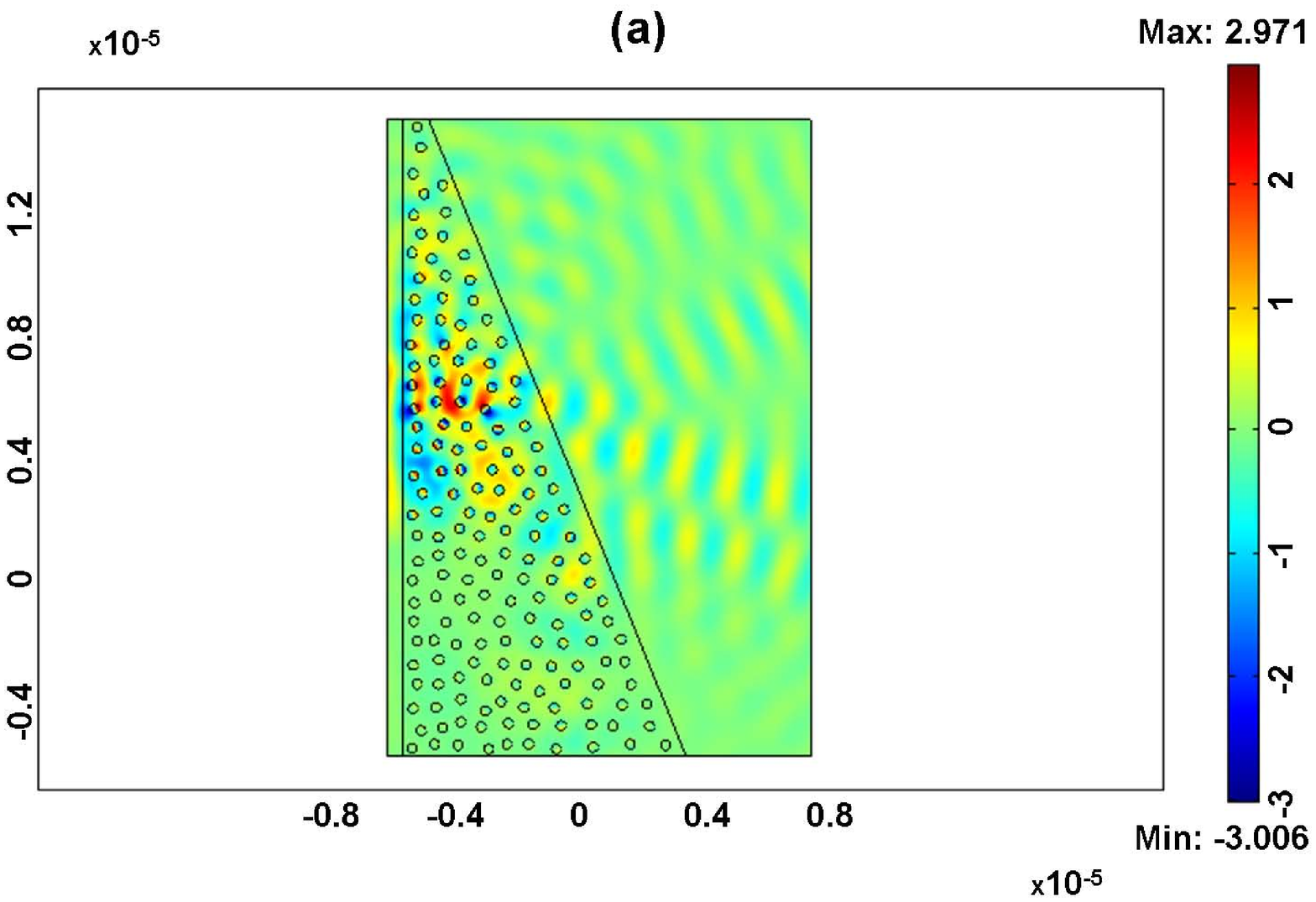}
\end{minipage}
\begin{minipage}{.49\linewidth}
\centering
\includegraphics[width=7cm]{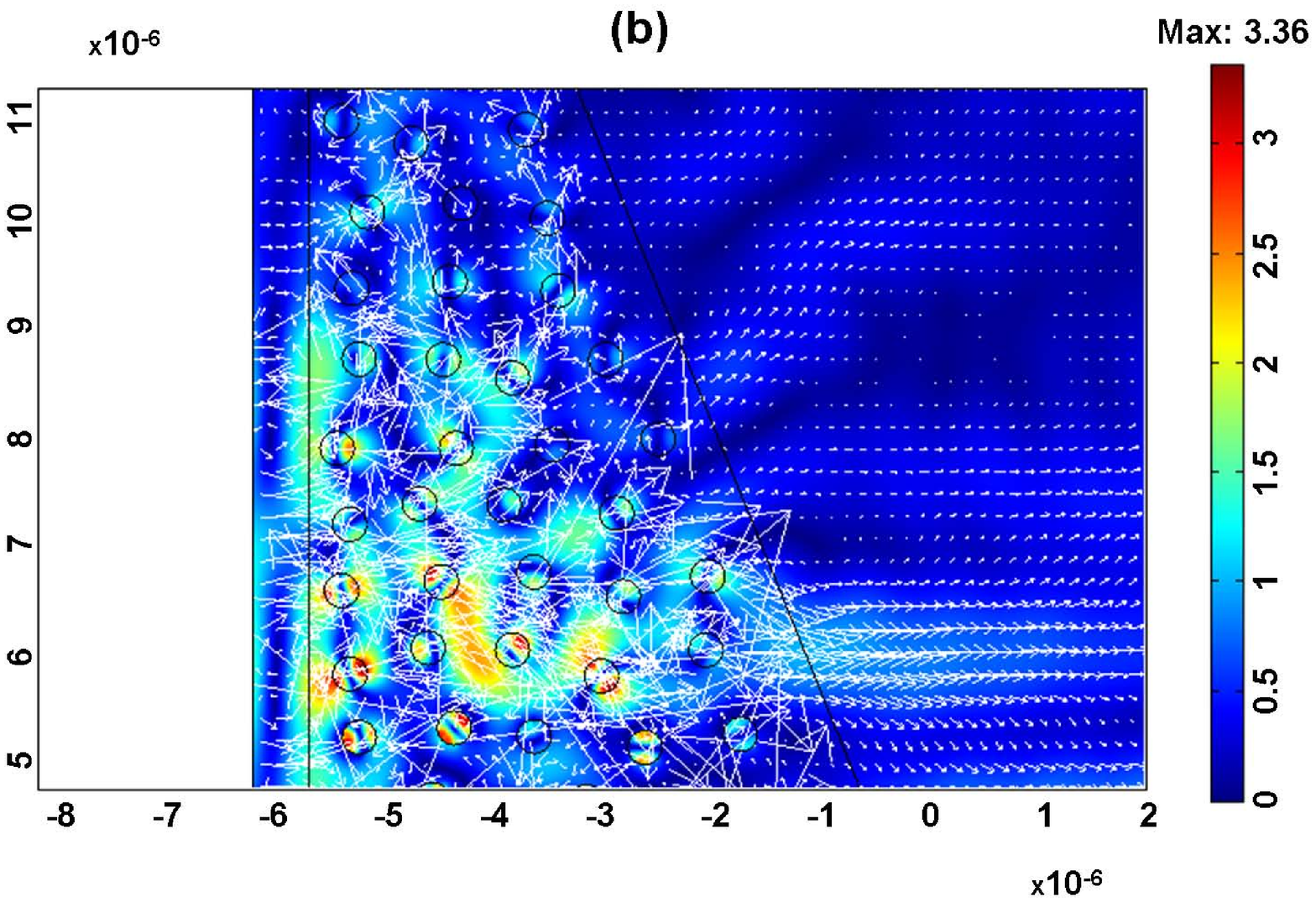}
\end{minipage}
\begin{minipage}{.98\linewidth}
\centering
\includegraphics[width=7cm]{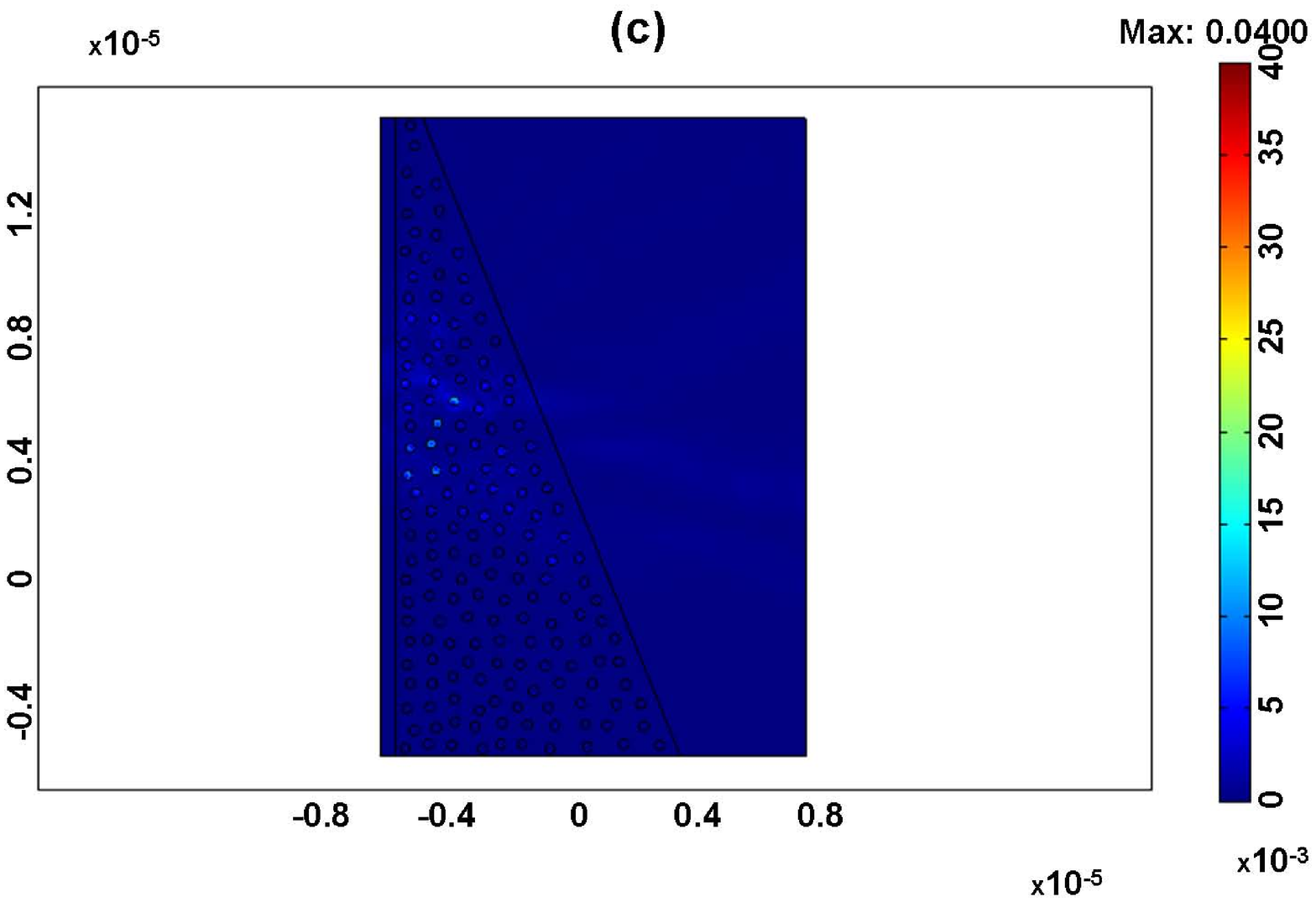}
\end{minipage}
\caption{ (a) Electric field $E_z({\bf R})$ in the configuration
shown in Figs. 9(a) - (c), (now with a top angle $\alpha=
22^\circ$), the Si rod array being now randomized but keeping the
same filling fraction $f= 0.30$ as in the ordered array. The
conditions of illumination are the same as in Figs. 8(a) - 8(c). (b)
Electric field norm $|E_z({\bf R})|$ (colors) and ${\bf <S(R)>}$
(arrows) in a detail of the prism upper area. (c) Averaged energy
flow norm ${\bf |<S(R)>|}$.}
\end{figure}

This latter remark is further confirmed by disordering this Si
cylinder array within the prism, (see Figs. 10(a) - (c)). Again, no
refracted beam into the air is observed, and a large portion of
energy is lost by scattering from the composite elements. Many
outgoing beams mix with each other, so that no forwardly transmitted
beam inside the prism, characterized by $<E_{z}({\bf R})>$ can be
distinguished by averaging over several realizations of the random
array. The apparent refracted beam at $\theta_t= 0^\circ$ shown in
the air side exiting the prism, changes as one varies the random
realization of rods.

%\newpage

\section{Conclusions}

%After proofreading the manuscript, tar and gzip the \texttt{.tex}
%file and figures; then enter the requested information into the
%\textit{Optics Express} online submission system at
%\url{http://www.opticsexpress.org} and upload the tarred and gzipped
%archive. If there is video or other multimedia, the associated files
%should be uploaded separately.

In this paper we have assessed the transmittance of composites of
dielectric particles whose Mie resonance electric and magnetic modes
were proposed by previous extensive studies to yield negative
refraction. We have shown that such structures cannot be homogenized
neither reproduce the propagation observed in the frequency regions
of study. In this way, we have proved that the negative refraction
previously found in ordered arrays is thus a diffraction effect
which disappears as soon as the particle distribution is randomized,
and does not reproduce the transmission of their corresponding EMT
uniform media, this was an open question so far.

In addition, we conclude, first, that the effective parameters
obtained from a homogenization theory, do not reproduce the
propagation observed through an ordered array of these elements.
This is further seen when these arrays are randomized. If an EMT
worked, it should not depend on whether the \lq\lq meta - atom"
distributions were ordered or disordered. Then strong scattering by
disordered rods extinguishes most of the incident energy, and there
is not negatively refracted forward beam observed.

Second, the behavior of the transmitted beam in ordered arrays also
depends on the shape of the sample and hence does not match with an
EMT which does not include this shape. These conclusions are
consistent with the well known {\it difficulty of working out EMTs
within the frequency range of resonance of the composite elements}
\cite{Zhang2011}. Similar consequences should hold for 3D composites
of resonant dielectric spheres.

\section*{Acknowledgements}

We thank J. J. S\'aenz and L. Froufe for stimulating and useful
discussions on this subject. Research supported by the Spanish
MiCINN through FIS2009-13430-C02-C01 and Consolider NanoLight
(CSD2007-00046) research contracts. The latter grant supports the
work of FJVV.



%% References





\end{document}